\documentclass[epsf,prb,twocolumn,showpacs,longbibliography,nofootinbib, superscriptaddress]{revtex4-1}

\usepackage[dvipsnames]{xcolor}
\definecolor{darkblue}{rgb}{0,0,.65}

\usepackage{microtype}
\usepackage{graphicx}
\usepackage[export]{adjustbox}
\usepackage{pifont}

\usepackage[
    pdfauthor   = {Ophelia Sommer},
    pdfstartview= FitH,
    breaklinks  = true,
    bookmarks   = true,
    colorlinks  = true,
    anchorcolor = black,
    citecolor   = blue,
    filecolor   = red,
    menucolor   = black,
    urlcolor    = blue,
    linkcolor   = blue,
]{hyperref}
\usepackage[all]{hypcap}

\usepackage{natbib}

\usepackage{amsmath,amssymb,amsthm}
\usepackage{siunitx}
\usepackage{braket} 
\usepackage{bm}
\usepackage{mathrsfs}
\usepackage{booktabs}

\makeatletter
\newcommand\notsotiny{\@setfontsize\notsotiny{6.5}{7.6}}

\usepackage{tikz}
\usetikzlibrary{decorations.markings}
\usetikzlibrary{calc,backgrounds,decorations.pathmorphing,decorations.pathreplacing}
\pgfdeclarelayer{background}
\pgfdeclarelayer{foreground}
\pgfdeclarelayer{super-foreground}
\usetikzlibrary{arrows}

\pgfsetlayers{background,main,foreground,super-foreground}

\tikzstyle{tensor}=[rectangle,rounded corners=2,draw=black]
\tikzstyle{tensorcirc}=[circle,draw=black]
\newcommand{\Z}{\mathbb{Z}}
\newcommand{\R}{\mathbb{R}}

\newcommand{\e}{\mathrm{e}}

\let\oldi\i
\renewcommand{\i}{\mathrm{i}}
\newcommand{\dd}{\mathrm{d}}

\newcommand{\mc}[1]{\mathcal{#1}}
\newcommand{\mr}[1]{\mathrm{#1}}
\newcommand{\bb}[1]{\mathbb{#1}}

\DeclareMathOperator{\tr}{Tr}
\let\Im\relax
\DeclareMathOperator{\Im}{Im}

\newcommand{\vb}[1]{\mathbf{#1}}

\theoremstyle{definition}

\def\be{\begin{equation}}
\def\ee{\end{equation}}

\begin{document}
\title{Higher Berry Curvature from the Wave function II:\\
	Locally Parameterized States Beyond One Dimension}

\author{Ophelia Evelyn Sommer}
\affiliation{Department of Physics, Harvard University, Cambridge MA 02138}
\author{Ashvin Vishwanath}
\affiliation{Department of Physics, Harvard University, Cambridge MA 02138}

\author{Xueda Wen}
\affiliation{School of Physics, Georgia Institute of Technology, Atlanta, GA 30332, USA}
\affiliation{Department of Physics, Harvard University, Cambridge MA 02138}
\affiliation{Department of Physics, University of Colorado, Boulder, CO 80309, USA}

\date{\today}
\begin{abstract}
	We propose a systematic wave function based approach to construct topological invariants
	for families of lattice systems that are short-range entangled
	using local parameter spaces. This construction is particularly suitable when given a
	family of tensor networks that can be viewed as the ground states of $d$ dimensional
	lattice systems, for which we construct the closed $(d+2)$-form higher Berry curvature,
	which is a generalization of the well known 2-form Berry curvature.
	Such $(d+2)$-form higher Berry curvature characterizes a flow of $(d+1)$-form higher
	Berry curvature in the system. Our construction is equally suitable for
	constructing other higher pumps, such as the (higher) Thouless pump in the
	presence of a global on-site $\mr{U}(1)$ symmetry, which corresponds to a
	closed $d$-form.
	The cohomology classes of such
	higher differential forms are topological invariants and are expected to be
	quantized for short-range entangled states.
	We illustrate our construction with exactly solvable lattice models that are in nontrivial higher Berry classes in $d=2$.
\end{abstract}

\maketitle
\tableofcontents
\section{Introduction and results}
\subsection{Introduction}
The Berry phase, which arises as a generic feature of the evolution of a quantum state,
has permeated all branches of physics ever since its discovery. As Berry showed\cite{berry84},
the Berry phase along a loop in the parameter space can be evaluated by integrating
the $2$-form Berry curvature over a surface enclosed this loop.
The integral of Berry curvature over a closed manifold is known to give rise to a quantized
topological invariant, the Chern number, which plays an essential role
in many topological phenomena such as the integer quantum Hall effect\cite{thoulessQuantizedHallConductance1982a}.

Recently, initiated by Kitaev's proposal\cite{kitaev2019},
the many-body generalization of Berry curvature of parameterized quantum systems has seen rapid developments.
\cite{ KS2020_higherberry, KS2020_higherthouless, Hsin_2020, Cordova_2020_i, Cordova_2020_ii, Else_2021,Choi_Ohmori_2022, Aasen_2022, qpump2022,
	Hsin_2023,Kapustin2201, Shiozaki_2022, 2022aBachmann,Ohyama_2022, ohyama2023discrete, Kapustin2305, homotopical2023, 2023Ryu, 2023Qi, 2023Shiozaki,2023Spodyneiko,2023Debray}.
In \onlinecite{kitaev2019},  Kitaev
outlined a construction of this higher Berry curvatures for families of Euclidean lattice systems
in general dimensions. The main motivation is to use these higher Berry curvatures and the
corresponding topological invariants to probe the topology of space of gapped
systems\cite{kitaevSimonsCenter1,kitaevSimonsCenter2,kitaevIPAM}. Later in \onlinecite{KS2020_higherberry},
based on a parameterized family of gapped Hamiltonians, Kapustin and Spodyneiko
gave an explicit construction of higher Berry curvatures in general dimensions
-- the Berry curvature $2$-form was generalized to higher Berry curvature $(d+2)$-form
for gapped systems in $d$ spatial dimensions.

Physically, such $(d+2)$ form higher Berry curvatures characterize a pump of $(d+1)$ form
higher Berry curvatures\cite{qpump2022}. For gapped systems, the integral of the higher
Berry curvature over a $(d+2)$ dimensional parameter space $X$ gives an invariant,
the so-called higher Berry invariant, that is believed to be quantized for short range
entangled systems and take values in $H^{d+2}(X,\mathbb Z)$, generalizing the Chern number
to gapped systems in $d\geq 1$ dimensions. When there is a global continuous symmetry $G$,
a similar construction based on the parameterized Hamiltonians gives rise to invariants of gapped
symmetric systems in $d\geq1$ dimensions\cite{KS2020_higherthouless}.
For example, in the case of $G=\mr{U}(1)$, one can generalize the well-known Thouless pump in $d=1$\cite{thouless_1983}
to higher Thouless pumps in $d>1$\cite{KS2020_higherthouless}.
\\

The higher Berry curvatures and higher Thouless pumps were studied systematically
in the framework of operator algebras\cite{Kapustin2201}.
Although one needs a parent Hamiltonian to define the higher Berry curvature
it was shown that the higher Berry invariants depend only on the family of ground states,
but not the choice of local parent Hamiltonians.

A natural question can be asked: Can one define the higher Berry curvature directly
from the wave function without using invoking a parent Hamiltonian? In \onlinecite{2024Sommer} we answer this question using Matrix product states in $d=1$ and in this work we, we answer the question by providing a systematic way of constructing
the higher Berry curvatures for locally parameterized states, such as tensor network states where each tensor can be locally parameterized\cite{2020RMP},
in general dimensions. 
It is known that tensor network states can represent the ground states
of a large class of quantum many-body systems. For example, in $d=1$,
any ground state of a gapped quantum spin chain can be represented efficiently using the
matrix product states (MPS) -- this has led to
the classification of all possible symmetry protected topological phases
for $d=1$ quantum spin systems\cite{Chen2011,Pollmann2012,Schuch2011}.
In this work, based on the family of locally parameterized states, we give an explicit construction of the $(d+2)$ form higher Berry curvature in a $d$ dimensional system.
When there is a global $\mr{U}(1)$ symmetry, we also construct the closed $d$-form
in a $d$ dimensional system, which characterizes the higher Thouless pump\cite{KS2020_higherthouless}.

By considering jointly the space of differential forms and simplex chains on a lattice, the expanded continuity equation for higher Berry curvature becomes a collection of descent equations relating a $p$ chain $q$ form to a $p+1$ chain $q+1$ form, which also appear in \onlinecite{kitaev2019,KS2020_higherberry}. Our novel technical insight is that these can be solved explicitly for arbitrary local parameter spaces, which means that we can define the higher Berry curvature and higher Thouless pumps in arbitrary dimensions simply in terms of the wavefunction and these local parameter spaces.  The explicit solution using Hamiltonian densities found in \onlinecite{KS2020_higherberry} is a special case of this construction. 

We also want to highlight recent efforts in extracting the higher Berry invariant from $d=1$
translationally-invariant or uniform
matrix product states (uMPS).
In \onlinecite{2023Ryu} and \onlinecite{2023Qi}, the authors study the higher Berry invariant by analyzing the gerbe structure of the parameterized family of $d=1$ uMPS, which was used in \onlinecite{2023Shiozaki}, to numerically calculate the associated invariant. The relation between these structures and our construction is explored in \cite{2024Sommer}. It would be interesting to see how such structures can be generalized to the non-translationally invariant and higher dimensional wave function ans\"atze considered here.
\subsection{Main results}

In this paper, we find a novel wavefunction based construction for the $(d+2)$-form higher Berry curvatures in $d$ dimensional systems, as well as the $d$-form for higher Thouless pumps in $d$ dimensional systems with a global $\mr{U}(1)$ symmetry, using local parameter spaces. See Eqs.\eqref{eq:HB_transport}, \eqref{eq:HB_curvature}, and Eqs.\eqref{eq:thouless_charge_flow}, \eqref{eq:d-form}, respectively.
Our results can be straightfowardly applied to tensor network states including the $d=1$ MPS results in Ref.\onlinecite{2024Sommer}.

\medskip 
Our approach to construct higher Berry curvatures and higher Thouless pumps requires the following data:
\begin{enumerate}
	\item A coarse graining of a spatial manifold $M$ called the lattice $\Lambda \subset M$.
	\item A collection of parameter spaces $X_p$ associated to each point $p\in\Lambda$, such that the total parameter space is a subset of their product $X\subseteq \prod_p X_p$. We only care about the germ of the parameter space inside this product space.
	\item A local fluctuation in a parameter space has an exponentially decaying impact on the observables of interest.
\end{enumerate}
If $q^{(n)}$ corresponds to the $n$-form (higher) charge of a conserved quantity,
we can construct the associated (higher) flow, and hence for non-compact $M$,
we find topological invariants associated with the transport of the (higher) charge to boundaries at infinity.
In particular, a many-body Hilbert space with a local factorisation $\mc{H}=\otimes_p\mc{H}_p$
and an associated family of gapped ground states can furnish such data in different ways.
By taking the local observables to be the $2$-form local Berry curvature $F^{(2)}$,
we construct the higher Berry curvature, while if the local observable is a $0$-form $\mr{U}(1)$ charge, we obtain the (higher) Thouless pump.

The rest of this paper is organized as follows:
In Sec.\ref{sec:MPSreview}, we give a brief review of the construction of higher Berry curvature from $d=1$ MPS,
and introduce the bulk-boundary correspondence that will be useful in studying higher dimensional systems.
In Sec.\ref{Sec:GeneralConstruct}, we generalize our construction to higher Berry curvatures and higher Thouless pumps in arbitrary dimensions.
In Sec.\ref{Sec:higherD},
we illustrate our construction with $d=2$ examples.
Then we conclude and discuss several future directions in Sec.\ref{sec:Discuss}.
There are also several appendices.
In appendix \ref{Sec:Unitaries} we illustrate that our approach is not intrinsic to tensor network states, or Hamiltonians. In particular we describe how the exactly solvable model considered in Sec.\ref{sec:MPSreview} gives the same higher Berry curvature when parameterised by a family of local unitaries.
In appendix \ref{app:KS},
we discuss the connection between the present approach and that in Kapustin and Spodeneiko\cite{KS2020_higherberry}, which can be viewed as choosing the local parameter space to be the space of coupling constants. 
We also give details on higher Thouless pumps in exactly solvable lattice models in appendix \ref{Sec:ThoulessPump}.

\section{Higher Berry curvatures from matrix product states}
\label{sec:MPSreview}

Before we introduce our construction of higher Berry curvatures in general dimensions, it is helpful to give a brief review of the construction in one dimension \cite{2024Sommer} to be self contained, and since the physical picture is more transparent. More details can be found in Ref.\onlinecite{2024Sommer}.

\subsection{Brief review of bulk formula for higher Berry curvatures from MPS}

Higher Berry curvature is associated with the flow of Berry curvature between boundaries at infinity in a lattice system\cite{qpump2022}. 
In Ref.\onlinecite{2024Sommer}, we make this connection clearly 
at the level of the wave function in one dimensional systems, by utilizing local parameter spaces, which is to say, that a family of wave functions will often have a notion of a local variation. We briefly review the main results of Ref. \onlinecite{2024Sommer} here.

To build up to the $d=1$ system of an infinite length, consider a finite subset of the one-dimensional lattice,
consisting of $N$ sites, such that the total Hilbert space $\mc{H}=\otimes_{p=1}^N \mc{H}_p$, 
where $\mc{H}_p$ is the local Hilbert space at site $p$ with a basis $\ket{s_p}$.
We consider a parameterized family of canonical matrix product states which can be viewed as the ground states of gapped systems:
\begin{align}
	\ket{\psi} & =\sum_{\set{s_p}} A_1^{s_1}\cdots A_N^{s_{N}}\ket{s_1\cdots s_N} \\
	           & =
	\begin{tikzpicture}[
			baseline ={([yshift=9]current bounding box.center)},
			inner sep=1mm]
		\node[tensor] (top) at (0,0.5) {$A_1$};
		\draw[-] (top)-- (top.east)-- +(0.3,0);
		\draw[-] (top.south)+(0,-0.3)--(top);
		\node[tensor] (top) at (1.0,0.5) {$A_2$};
		\draw[-] (top.west)+(-0.3,0)--(top)-- (top.east)-- +(0.3,0);
		\draw[-] (top.south)+(0,-0.3)--(top);
		\node[tensor] (top) at (3.0,0.5) {$A_N$};
		\draw[-] (top.west)+(-0.3,0)--(top);
		\draw[-] (top.south)+(0,-0.3)--(top);
		\node at (0,-0.3) {$s_1$};
		\node at (1,-0.3) {$s_2$};
		\node at (3,-0.3) {$s_N$};
		\node at (2,0.5) {$\cdots$};
		\node at (2,-0.3) {$\cdots$};
	\end{tikzpicture}
\end{align}
where we have used the conventional tensor network contraction diagrams (The reader is directed towards the reviews \onlinecite{2020RMP} and \onlinecite{ORUS2014117} for details on MPS). We consider this family of MPS to be functions $X\to \mc{H}$, with associated exterior derivatives $\dd=\sum_p \dd_p$. Here the local derivative $\dd_p$ is just to vary the parameters in the tensor at site $p$.

The total Berry curvature for $|\psi\rangle$ is the well known differential two-form
$\Omega^{(2)}=-\Im\braket{\dd\psi|\dd\psi}=\dd \mc{A}$, where
$\mc{A}=-\Im\braket{\psi|\dd \psi}$ is the one-form Berry connection, and we always take the state $\ket{\psi}$ to be normalized.
As discussed in Ref.\onlinecite{2024Sommer}, we proposed to take $F_p^{(2)}=\dd_p\mc{A}$ as the definition of the Berry curvature at site $p$, and it is obvious that $\Omega^{(2)}=\sum_p F_p^{(2)}$.
In general, because there is entanglement between site $p$ and the rest of the system, $F_p^{(2)}$ is not a closed two-form. It needs not have a quantized integral over a closed 2-manifold of parameters $X$, i.e., $\int_X F^{(2)}_p\notin 2\pi\Z$, and further consider deforming $X$ into some other two manifold $Y$, then $\int_X F_p^{(2)}\neq \int_Y F_p^{(2)}$. In particular if $Z$ is a $3$-manifold with $X$ and $Y$ as boundaries, then $\int_{X-Y}F_p^{(2)}=\int_Z \dd F_p^{(2)}\neq 0$ by Stokes theorem. Thus the Berry curvature of a point can change, but since the total Berry curvature $\sum_p F_p^{(2)}$ is closed, the way for the Berry curvature to change at a point is if it flows from another. Thus we define the flow $p\to q$ of Berry curvature as $F^{(3)}_{pq}$, which satisfies the continuity equation 
\begin{equation}
    \sum_p F^{(3)}_{pq}=\dd F_{q}^{(2)}.\label{eq:continuity1d}
\end{equation} 
Using the local variation, a solution to this continuity equation is 
\be
\label{eq:F3}
F_{pq}^{(3)}=\dd_p F_q^{(2)}=\dd_p\dd_q \mathcal A.
\ee
Any other solution is related to this one by the addition of a flow that is a pure circulation
- that is if $\tilde{F}_{pq}^{(3)}$ is a solution, then
$\tilde{F}_{pq}^{(3)}-F^{(3)}_{pq}=\sum_r C_{pqr}$, for some three-form $C_{pqr}$ which
is completely anti symmetric in $p,q,r$. This ambiguity does not modify the invariants we are interested in.

Next, to define the flow of Berry curvature from the left boundary to the right boundary, we should pick a middle of the system say some point $a$, then the higher Berry curvature which characterizes this flow can be written as
\be
\label{Omega3}
\Omega^{(3)}=\sum_{p<a<q}F_{pq}^{(3)}.
\ee
For a gapped state, it is expected that $F_{pq}^{(3)}$ decays exponentially as a function of the distance $|p-q|$.
From this point of view, $\Omega^{(3)}$ can be viewed as a local quantity near $x=a$ that captures the flow of Berry curvature, and the definition in \eqref{Omega3} works well for an infinite $1d$ system.
Because of this local property, $\Omega^{(3)}$ may be different depending on the choice of $a$.
However, the integral of $\Omega^{(3)}$ over a closed 3-manifold $X$ is a quantized topological invariant, i.e., 
\be
\int_X \Omega^{(3)}\in 2\pi \mathbb Z.
\ee
This quantization property has been proved for $d=1$ uMPS in Ref.\onlinecite{2024Sommer}. The underlying physics of this topological invariant corresponds to the Chern number pump\cite{qpump2022}, which is an analogy of the well known Thouless charge pump in $1d$.
As illustrated in Ref.\onlinecite{2024Sommer}, the formula in \eqref{Omega3} in terms of MPS is suitable for both analytic and numerical calculations in a general lattice system.

It is emphasized that the above construction is not limited to a tensor network state, where one can vary the parameters in a local tensor. One can consider more general states that are locally parameterized. See, e.g., a concrete example in appendix \ref{Sec:Unitaries}, where the family of states are related by local unitary transformations.

Furthermore,  the Higher Berry curvature $\Omega^{(3)}$  in \eqref{Omega3} is related to the Schmidt decomposition across a cut between regions $A$ and $B$, $\ket{\psi}=\sum_{\alpha}c_\alpha\ket{\alpha}_{[A]}\ket{\alpha}_{[B]}$. Using left canonical MPS it takes the simple form\cite{2024Sommer}
\begin{equation}\Omega^{(3)}=\Im\sum_{\alpha}\dd c_\alpha^2\braket{\dd\alpha|\dd\alpha}_{[A]},
\end{equation} 
where the tensor network provides a regularisation scheme to make sense of this formula that naively consists of infrared divergent terms\cite{2024Sommer}.

\subsection{Bulk boundary correspondence}

 We can also study the higher Berry invariant via the bulk boundary correspondence, as in Ref.\onlinecite{qpump2022}. 
This discussion will be useful for our later study of higher dimensional systems. Here in $d=1$, by introducing a fictitious boundary, the higher Berry invariant will equal the accumulated Chern number of this boundary.

To illustrate this bulk-boundary correspondence, we consider the exactly solvable model as studied recently in Refs.\onlinecite{qpump2022,2023Qi,2024Sommer}.
Let $\Lambda=\mathbb{Z}$ be a $d=1$ lattice of spin $1/2$ systems, with local Hilbert spaces $\mc{H}_p=\mathbb{C}^2$. We associate to each site Pauli matrices, and let $\ket{\vb n}$ be a coherent state along the direction of $\vb n\in \R^3$. Consider the parameter space $X=S^3$ embedded into $\R^{4}$ as $(\vb w,w_4)$ satisfying $\vb w^2+w_4^2=1$, where $\vb w=(w_1,w_2,w_3)$. While we do not require a Hamiltonian to compute the higher Berry curvature, to explain how we arrive at our particular choice of wave function, we consider the ground state of a nearest neighbour Hamiltonian
\begin{equation}
	H_\mr{1d}(w)=\sum_p H_{p}^\mr{onsite}(w)+H_{p,p+1}^\mr{int}(w),
	\label{eq:model_ham}
\end{equation}
where the onsite term is a single-spin term that takes the form of a Zeeman coupling with alternating sign
\begin{equation}
	\label{H_1body}
	H_p^\mr{onsite}(w)=(-1)^p\vb w\cdot \bm{\sigma}_p,
\end{equation}
and the interaction term is a two-spin term whose coefficient depends on $w_4$:
\begin{equation}
	H_{p,p+1}^\mr{int}=g_p(w_4)\bm{\sigma}_p\cdot \bm{\sigma}_{p+1}.
\end{equation}
The coupling constant takes the form
\begin{equation}
	g_p(w_4)=\begin{cases}
		w_4,  & p\in 2\Z+1 \text{ and } w_4>0 \\
		-w_4, & p\in 2\Z \text{ and } w_4<0   \\
		0,    & \text{otherwise}
	\end{cases}.
	\label{eq:gp}
\end{equation}
Note that $\vb w^2=1-w_4^2$, so the onsite coupling vanishes at the poles $w_4=\pm1$. Here the ground state forms one of the two spin singlet coverings of $\Z$ (depending on which pole), and for all parameters the Hamiltonian completely dimerizes and is gapped.
The Hamiltonian can be visualized for different values of $w_4 \in [-1,1]$ as:
\begin{equation}
	\label{eq:H_config}
	\small
	\begin{tikzpicture}

		\node at (-60pt,0pt){$0<w_4<1$:};

		\draw [dashed](-10+20pt,30pt)--(-10+20pt,-76pt);

		\notsotiny
		\draw (-20pt,0pt) circle (4.5pt);
		\node at (-20pt,0pt){$+$};
		\draw [thick](4.5-35pt,0pt)--(15.5-40pt,0pt);

		\draw (0pt,0pt) circle (4.5pt);
		\node at (0pt,0pt){$-$};
		\draw [thick](4.5pt,0pt)--(15.5pt,0pt);
		\draw (20pt,0pt) circle (4.5pt);
		\node at (20pt,0pt){$+$};

		\draw (40pt,0pt) circle (4.5pt);
		\node at (40pt,0pt){$-$};
		\draw [thick](44.5pt,0pt)--(55.5pt,0pt);
		\draw (60pt,0pt) circle (4.5pt);
		\node at (60pt,0pt){$+$};

		\draw (80pt,0pt) circle (4.5pt);
		\node at (80pt,0pt){$-$};
		\draw [thick](84.5pt,0pt)--(95.5pt,0pt);
		\draw (100pt,0pt) circle (4.5pt);
		\node at (100pt,0pt){$+$};

		\draw (120pt,0pt) circle (4.5pt);
		\node at (120pt,0pt){$-$};
		\draw [thick](124.5pt,0pt)--(135.5pt,0pt);
		\draw (140pt,0pt) circle (4.5pt);
		\node at (140pt,0pt){$+$};

		\begin{scope}[yshift=-22pt]
			\small
			\node at (-51pt,0pt){$w_4=0$:};

			\notsotiny
			\draw (-20pt,0pt) circle (4.5pt);
			\node at (-20pt,0pt){$+$};

			\draw (0pt,0pt) circle (4.5pt);
			\node at (0pt,0pt){$-$};

			\draw (20pt,0pt) circle (4.5pt);
			\node at (20pt,0pt){$+$};

			\draw (40pt,0pt) circle (4.5pt);
			\node at (40pt,0pt){$-$};

			\draw (60pt,0pt) circle (4.5pt);
			\node at (60pt,0pt){$+$};

			\draw (80pt,0pt) circle (4.5pt);
			\node at (80pt,0pt){$-$};

			\draw (100pt,0pt) circle (4.5pt);
			\node at (100pt,0pt){$+$};

			\draw (120pt,0pt) circle (4.5pt);
			\node at (120pt,0pt){$-$};

			\draw (140pt,0pt) circle (4.5pt);
			\node at (140pt,0pt){$+$};

		\end{scope}

		\begin{scope}[yshift=-44pt]

			\small

			\node at (-63pt,0pt){$-1<w_4<0$:};

			\notsotiny

			\draw (-20pt,0pt) circle (4.5pt);
			\node at (-20pt,0pt){$+$};

			\draw (0pt,0pt) circle (4.5pt);
			\node at (0pt,0pt){$-$};
			\draw [thick](4.5-20pt,0pt)--(15.5-20pt,0pt);
			\draw (20pt,0pt) circle (4.5pt);
			\node at (20pt,0pt){$+$};

			\draw (40pt,0pt) circle (4.5pt);
			\node at (40pt,0pt){$-$};
			\draw [thick](24.5pt,0pt)--(35.5pt,0pt);
			\draw (60pt,0pt) circle (4.5pt);
			\node at (60pt,0pt){$+$};

			\draw (80pt,0pt) circle (4.5pt);
			\node at (80pt,0pt){$-$};
			\draw [thick](64.5pt,0pt)--(75.5pt,0pt);
			\draw (100pt,0pt) circle (4.5pt);
			\node at (100pt,0pt){$+$};

			\draw (120pt,0pt) circle (4.5pt);
			\node at (120pt,0pt){$-$};
			\draw [thick](104.5pt,0pt)--(115.5pt,0pt);
			\draw (140pt,0pt) circle (4.5pt);
			\node at (140pt,0pt){$+$};
			\draw [thick](144.5pt,0pt)--(150.5pt,0pt);
		\end{scope}

		\begin{scope}[yshift=-66pt]

			\draw [thick](-20pt,0pt)--(0pt,0pt);
			\draw [thick](20pt,0pt)--(40pt,0pt);
			\draw [thick](60pt,0pt)--(80pt,0pt);
			\draw [thick](100pt,0pt)--(120pt,0pt);
			\draw [thick](140pt,0pt)--(150pt,0pt);

			\node at (140pt,0pt){$\bullet$};
			\node at (120pt,0pt){$\bullet$};
			\node at (100pt,0pt){$\bullet$};
			\node at (80pt,0pt){$\bullet$};
			\node at (60pt,0pt){$\bullet$};
			\node at (40pt,0pt){$\bullet$};
			\node at (20pt,0pt){$\bullet$};
			\node at (0pt,0pt){$\bullet$};
			\node at (-20pt,0pt){$\bullet$};

			\small
			\node at (-55pt,0pt){$w_4=-1$:};
		\end{scope}

		\begin{scope}[yshift=22pt]

			\draw [thick](-30pt,0pt)--(-20pt,0pt);
			\draw [thick](0pt,0pt)--(20pt,0pt);
			\draw [thick](40pt,0pt)--(60pt,0pt);
			\draw [thick](80pt,0pt)--(100pt,0pt);
			\draw [thick](120pt,0pt)--(140pt,0pt);

			\node at (140pt,0pt){$\bullet$};
			\node at (120pt,0pt){$\bullet$};
			\node at (100pt,0pt){$\bullet$};
			\node at (80pt,0pt){$\bullet$};
			\node at (60pt,0pt){$\bullet$};
			\node at (40pt,0pt){$\bullet$};
			\node at (20pt,0pt){$\bullet$};
			\node at (0pt,0pt){$\bullet$};
			\node at (-20pt,0pt){$\bullet$};

			\small
			\node at (-52pt,0pt){$w_4=1$:};

		\end{scope}

	\end{tikzpicture}
\end{equation}
We use $\pm$ to represent the sign of the Zeeman coupling at this site, while $\bullet$ represents the case of vanishing Zeeman coupling. Interaction terms are represented by solid lines joining pairs of lattice sites. It is convenient to parameterise $X$ with  hyperspherical coordinates:
\begin{alignat*}{2}
	\label{HS_coordinate}
	 & w_1=\sin(\alpha)\sin(\theta)\cos(\phi), & \quad & w_2=\sin(\alpha)\sin(\theta)\sin(\phi), \\
	 & w_3=\sin(\alpha)\cos(\theta),           &       & w_4=\cos(\alpha),
\end{alignat*}
where $0\leq \alpha, \theta\leq \pi, \, 0\leq \phi\leq 2\pi$.
By applying the MPS formula of our higher Berry curvature in \eqref{Omega3} and choosing $x=a$ along the dashed line in \eqref{eq:H_config}, it was found that\cite{2024Sommer}
\be
\label{eq:3-form_toy}
\Omega^{(3)}=\frac12\cos\alpha\sin\theta~\dd\alpha\wedge\dd\theta\wedge\dd\phi,
\ee
for $0\le \alpha\le \pi/2$, and $\Omega^{(3)}=0$ otherwise. 
One can check explicitly that $\int_{X=S^3} \Omega^{(3)}=2\pi$.

\medskip

Now we introduce a physical boundary at a point $N\in 2\Z$, so that the system is defined on the lattice $\Z_{\leq N}$, and consider the same Hamiltonian \eqref{eq:model_ham}. The flow of Berry curvature $F_{pq}^{(3)}$ decays exponentially in the distance $|p-q|$. As such, taking a cut $a$ deep in the bulk, the higher Berry invariant defined on the whole lattice $\Z$  will differ exponentially little from the one defined on this truncated system. From equation \eqref{eq:continuity1d}, and noting that the edge defined as $a<p\leq N$ is finite, we can define
\begin{equation}
    \Omega^{(3)}=\dd\sum_{p>a}F^{(2)}_p=\dd \omega^{(2)}
\end{equation}
where $\omega^{(2)}=\sum_{p>a}F^{(2)}_p$ is the boundary berry curvature.
In the exactly solvable model, if $a\in 2\Z+1/2$, then the boundary is decoupled when $-1\leq w_4\leq 0$. Thus we find
\begin{equation}
	\omega^{(2)}=\begin{cases}
		\frac{1}{2}\sin\alpha\sin\theta~\dd\theta\wedge\dd\phi, & 0\leq w_4\leq 1,  \\
		\frac{1}{2}\sin\theta~\dd\theta\wedge \dd\phi,          & -1\leq w_4\leq 0,
	\end{cases}
\end{equation}
which is of course just the Schmidt weighted Berry curvature of the boundary. Because the higher Berry curvature of the model is nontrivial, $\omega^{(2)}$ cannot be well defined over the whole of $X=S^{3}$, which manifests as a gapless Weyl point at $w_4=-1$. If we deform the model by taking the boundary parameter space to exclude this Weyl point $X_\mr{bdy}=D^{(3)}=S^3\setminus\set{w_4=-1}$, the Berry invariant will be 
\begin{equation}
	\int_{S^3_\mr{bulk}}\Omega^{(3)}_\mr{bulk}=\int_{D^3}\Omega^{(3)}=\int_{S^2}\omega^{(2)}=2\pi
\end{equation}
where $S^{2}=S^{3}\cap\set{w_4=1+0^+}$. 
An interesting way of computing the Higher Berry curvature in general is by the clutching construction\cite{qpump2022}. It is not possible to have a fully gapped boundary for any particular truncation. However, considering $X$ as a union of contractible parameter spaces $U_\alpha$, we can truncate differently for each contractible space $U_\alpha$, and reconstruct the Berry invariant from the Berry curvature of the  intersections $U_\alpha\cap U_\beta$. In particular, for the model we can consider the two hemispheres $S^3=D_N^3\cup D_S^{3}$ where $D_N^{3}=\set{w\in S^3|w_4\geq 0}$ and $D_S=\set{w\in S^3|w_4\leq 0}$. Then the higher Berry invariant will correspond to the difference in the edge curvature $\omega_{N/S}^{(2)}$ on the equator $D_N^3\cap D_S^{3}=S^2$. As before, on the northern hemisphere we can truncate at site $N$, while for the southern hemisphere we truncate at site $N-1$. Pictorially the two MPS are
\begin{equation}
	\label{MPS_clutch}
	\begin{tikzpicture}[
			baseline ={([yshift=-7]current bounding box.center)},
			inner sep=1mm]
		\draw [dashed](-10+24.5pt,25pt)--(-10+24.5pt,-50pt);
		\node at (-2,0.5) {$D^3_N:$};
		\node at (-0.8,0.5) {$\cdots$};
		\node[tensor] (top) at (0,0.5) {$A$};
		\draw[-] (top.west)+(-0.3,0)--(top)-- (top.east)-- +(0.3,0);
		\draw[-] (top.south)+(0,-0.3)--(top);
		\node[tensor] (top) at (1.0,0.5) {$A$};
		\draw[-] (top.west)+(-0.3,0)--(top);
		\draw[-] (top.south)+(0,-0.3)--(top);
		\node[tensor] (top) at (2.5,0.5) {$A$};
		\draw[-] (top.west)+(-0.3,0)--(top)-- (top.east)-- +(0.3,0);
		\draw[-] (top.south)+(0,-0.3)--(top);
		\node[tensor] (top) at (3.5,0.5) {$A$};
		\draw[-] (top.west)+(-0.3,0)--(top);
		\draw[-] (top.south)+(0,-0.3)--(top);
		\node at (0,-0.2) {$s_p$};
		\node at (1,-0.2) {$s_{p+1}$};
		\node at (2.55,-0.2) {$s_{N-1}$};
		\node at (3.5,-0.2) {$s_N$};
		\node at (1.7,0.5) {$\cdots$};
		\node at (1.7,-0.2) {$\cdots$};
		\begin{scope}[yshift=-45pt]
			\node at (-2,0.5) {$D^3_S:$};
			\node at (-0.8,0.5) {$\cdots$};
			\node[tensor] (top) at (0,0.5) {$A$};
			\draw[-] (top.west)+(-0.3,0)--(top)-- (top.east)-- +(0.3,0);
			\draw[-] (top.south)+(0,-0.3)--(top);
			\node[tensor] (top) at (1.0,0.5) {$A$};
			\draw[-] (top.west)+(-0.3,0)--(top);
			\draw[-] (top.south)+(0,-0.3)--(top);
			\node[tensor] (top) at (2.5,0.5) {$A$};
			\draw[-] (top.west)+(-0.3,0)--(top);
			\draw[-] (top.south)+(0,-0.3)--(top);
			\node at (0,-0.2) {$s_p$};
			\node at (1,-0.2) {$s_{p+1}$};
			\node at (2.55,-0.2) {$s_{N-1}$};
			\node at (1.7,0.5) {$\cdots$};
			\node at (1.7,-0.2) {$\cdots$};
		\end{scope}
	\end{tikzpicture}
\end{equation}
Each MPS is a fully gapped state, and there is no obstruction to defining $\omega^{(2)}_{N/S}$ globally on each chart. In particular it is clear that 
$\omega_N^{(2)}=\sum_{a<p\leq N}F_p^{(2)}$ while $\omega^{(2)}_S=\sum_{a<p\leq N-1}F_p^{(2)}$ so on the equator their difference is just $\omega_N^{(2)}-\omega_S^{(2)}=F_N^{(2)}=\frac{1}{2}\sin\theta\dd\theta\wedge\dd \phi$. Then the higher Berry invariant can be calculated using Stokes theorem
\begin{equation}
    \int_{S^3}\Omega^{(3)}=\int_{D_N^{3}\cup{D_S^3}}\Omega^{(3)}=\int_{S^2}\omega_N^{(2)}-\omega_S^{(2)}=2\pi
\end{equation}
The 2-form $\omega^{(2)}$ can be considered a $2$-connection for the higher Berry curvature over the charts of the parameter space.

This bulk-boundary correspondence will be useful in studying the higher dimensional systems, as we will see later in Sec.\ref{Sec:higherD}.
\section{Higher Berry curvature and Thouless pump for locally parameterised states}
\label{Sec:GeneralConstruct}
In this section, we introduce a systematic way of constructing higher Berry curvatures and higher Thouless pumps from local parameter spaces in general dimensions with any notion of local parameter spaces.

\subsection{General formalism}
\label{Sec:general_formalism}

We consider a gapped system living on a spatial
$d$-dimensional lattice $\Lambda$,
which is a discrete subset of a spatial manifold $M$ with a metric $g$, such that
$\Lambda$ has no accumulation points, and there is a distance $r$ such that any
point in $M$ is at most a distance $r$ from some point in $\Lambda$.
For simplicity, one may take $M=\R^d$ and $\Lambda=\mathbb Z^d$, although we are not limited to this choice.
We associate a finite dimensional local Hilbert space to each point $p\in
	\Lambda$ which we denote $\mc{H}_p$, along with a manifold of
local parameters $X_p$. The total Hilbert space is the tensor product
$\mc{H}=\otimes_{p\in \Lambda} \mc{H}_p$, and the total parameter space is
a submanifold of the Cartesian product of local parameter spaces $X\subseteq\prod_{p\in \Lambda} X_p$. The gapped condition is intwined with the notion of local parameter space. In particular there is a correlation length $\xi$ such that for any local observable of interest $O_{p_0\cdots p_n}$ on sites $p_0\cdots p_n$ obeys some cluster decomposition with vanishing vacuum value, so that $O_{p_0\cdots p_n}=\mc{O}(\exp(-\mr{max}_{ij}g(p_i,p_j)/\xi))$ (note if $O_{p_0\cdots p_n}$ is a differential form, we imagine that this holds after integrating over the parameter manifold).
Then the local variations in parameters at any site $q$, which we denote with the exterior derivative $\dd_q$, will cause a change $\dd_q O_{p_0\cdots p_n}$ which we require to again have this finite correlation length. We could think of these local parameter spaces as considering a quantum as parameterized over not just over some global external variable, but rather a classical background field, whose response is gapped.

\begin{figure*}
	\centering

	\includegraphics[width=0.7\linewidth]{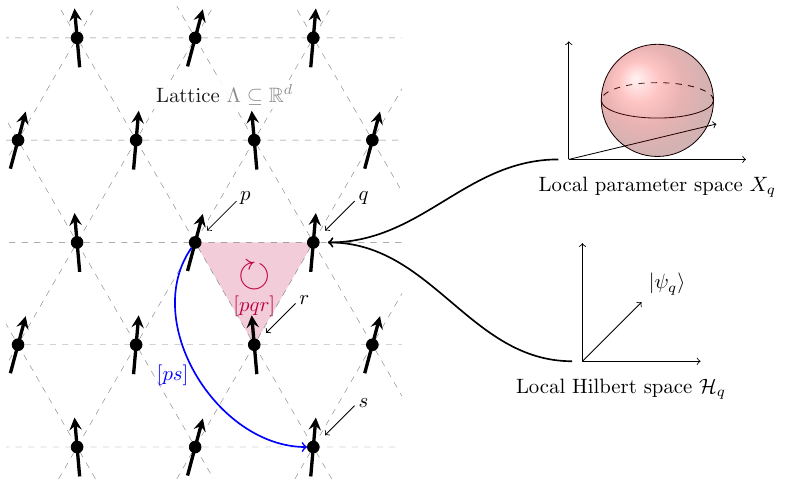}
	\caption{A locally parameterised system on a lattice $\Lambda$ means we associate to each $q\in\Lambda$ both a Hilbert space and a local space of parameters.
		Observables live on simplexes of the lattice, consisting of points $[p]$, flows $[ps]$ between points $p$ and $s$, circulations $[pqr]$ $p\to q\to r$ and so on. Note that we do not only consider simplexes which are nearest neighbours on a lattice.}
	\label{fig:construction_data}
\end{figure*}

We seek to keep track of two types of information, the spatial geometry of the
lattice $\Lambda$, and the geometry on the parameter space $X$. For the latter,
the notion of differential forms is well suited, as the parameter space can be taken to be smooth. We assume differential forms are familiar to the audience.
For the spatial geometry, we need to reckon with the discrete nature of
the lattice, and instead follow Refs.
\onlinecite{KS2020_higherberry,KS2020_higherthouless,Kapustin2201} and resort to the notion of oriented simplexes, see figure \ref{fig:construction_data} for an illustration of the data we require.
Briefly, a coarse $n$-simplex chain is the linear space which is spanned by all of the oriented convex hulls of $n+1$ points $p_0\cdots p_n$, denoted $[p_0\cdots p_n]$. To any simplex we assign a distance from the diagonal to be $\mathrm{dist}[p_0\cdots p_n]=\mathrm{max}_{ij}g(p_i,p_j)$. We are interested in chains with finite correlation length, where the coefficients of the chain $c_{p_0\cdots p_n}$ decay as $c_{p_0\cdots p_n}=\mc{O}\left(\exp(-\mr{dist}[p_0\cdots p_n]/\xi)\right)$, which are also called controlled\cite{roe2003lectures}. There is a boundary map $\partial$ mapping an $n$ chain to an $n-1$ chain, $\partial[p_0\cdots p_n]=\sum_l(-)^l[p_0\cdots\hat{p}_l\cdots p_n]$, such that $\partial^2=0$, so the chains give rise to a chain complex $\mc{C}_\bullet(\Lambda)$. The dual vector space is spanned by the dual or cosimplexes $(p_0\cdots p_n)$ and have coboundary $\delta$. A \emph{cocontrolled} cosimplex chain must have only a finite number of non-zero terms within any radius of the diagonal, and so the corresponding cochain complex $\mc{C}^\bullet(\Lambda)$ for any $\Lambda \subseteq \R^{d}$ is concentrated in the $d$'th degree, and is generated e.g. by the conical partition $b$ of $\Lambda$ into $d+1$ regions $\Lambda_i$, in particular $b=\sum_{p_i\in \Lambda_i}[p_0\cdots p_n]$\cite{Kapustin2201}.
We combine the differential forms on $X$ corresponding to a ``conserved'' observable $q^{(n)}_{p_0\cdots p_l}$, into finitely correlated simplex-chain (note that we use Einstein summation convention) $q^{(n)}=\frac{1}{(l+1)!}q^{(n)}_{p_0\cdots p_l}[p_0\cdots p_l]$ that satisfies $\dd \partial q^{(n)}=0$. The continuity equation relating the change in $q^{(n)}$ to the accumulation of a current $q^{(n+1)}$ becomes
\begin{equation}
	\partial q^{(n+1)}=\dd q^{(n)}.\label{eq:descent}
\end{equation}
This is the \emph{descent equation}, and by iterating we can find all \emph{higher charges} $q^{(>n)}$ related to $q^{(n)}$. By contracting any higher charge $q^{(m)}$ with a closed, non-exact cochain $b$, we find a closed non-exact differential form  $\braket{q^{(m)},b}$ on $X$ and therefore a topological invariant of the family of $d$-dimensional system when integrated. The choice of $b$ will shift the differential form by at most an exact form, and hence not modify the topological invariant.

\subsection{Higher Berry curvature and higher Thouless pump}
\label{subsec:solve_descent}

\begin{figure*}
	\centering
	\includegraphics[width=0.8\linewidth]{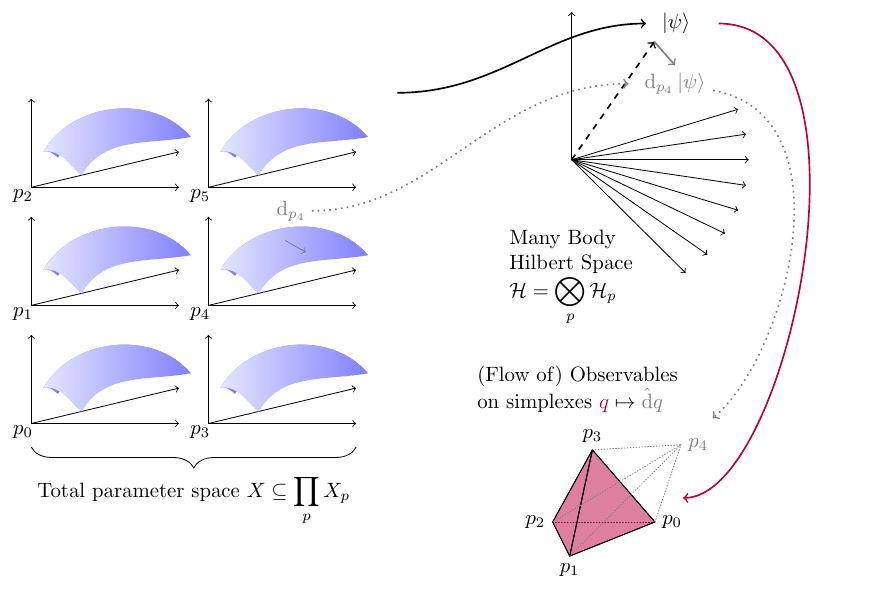}
	\caption{Graphical illustration of the action of $\hat\dd$ defined in equation \eqref{eq:dhat_def}. For each site in the lattice $\Lambda$ we have a parameter space $X_p$, and by specifying the parameter value at each point, which is itself a point in $X\subseteq \prod_p X_p$, we find a short-range entangled wave function. From the wave function we can calculate observables, whether they be local to a point in the lattice, measuring the flow between two such points, a circulation etc., such observables take values that are naturally associated with (linear combinations of) simplexes. The action of $\hat\dd$ is to consider the variation of this observable with respect to local parameters at all points. In particular, in this illustration we consider variations with respect to some point $p_4$. This changes the wave function, and hence changes the value of the observable $\dd_{p_4}q$. We associate the rate of change with an observable living on a new simplex, which we obtain by appending the point $p_4$.
	}
	\label{fig:dhat_illustration}
\end{figure*}
By solving the descent equations \eqref{eq:descent} we can obtain the flow of Berry curvature, and hence the higher Berry curvature, and this task is greatly facilitated by the use of local parameter spaces. Such a solution is
necessarily non-unique, since if $q^{(n+1)}$ solves it, then we may add the
boundary of any chain and obtain a new solution $q^{(n+1)}+\partial[\cdots]$.
Varying only the parameters at a
single site $p$ gives us a collection of exterior derivative operators
$\dd_p$, such that $\dd =\sum_{p\in \Lambda}\dd_p$. We would like these to act on the space of simplex chain valued differential
forms.  We define the simplex-exterior derivative
\begin{equation}
	\small
	\hat{\dd}:\frac{1}{(n+1)!}c_{p_0\cdots p_n}[p_0\cdots p_n]\mapsto\frac{1}{(n+1)!} \dd_q c_{p_0\cdots
			p_n}[qp_0\cdots p_n].\label{eq:dhat_def}
\end{equation}
A pictorial interpretation of this operation can be seen in figure \ref{fig:dhat_illustration}.
To understand this operation, consider the case of the charge $0$-chain $q=
	q_p[p]$. The simplex-exterior derivative is a current $\hat\dd
	q=\frac{1}{2}(\dd_p q_q-\dd_qq_p)[pq]$. This is exactly the solution to the question the descent equation
poses, if the total charge $Q=\sum_pq_p$ is conserved. Indeed we can verify that $\partial (\hat\dd q)=\dd q-\dd_p(Q)[p]$,
so if the total charge cannot be modified by local parameter variations, then
$\hat\dd q$ solves the descent equation of $q$. The simplex-exterior can be viewed as a
factorisation of the total exterior derivative $\dd =\partial \hat \dd+\hat\dd \partial$.
It is straightforward to check that
\begin{equation}
	\partial\left(\hat{\dd}^n q\right)=n\dd \hat\dd^{n-1}q+(-1)^n\hat\dd^n\partial q.
\end{equation}
Recall the descent equation \eqref{eq:descent}, it seems as if, ignoring this second term, it would be possible to solve the descent equation for $q^{(n)}$ simply by applying $\hat\dd$. A full solution requires us to think about this second term however, but it is most often the case that $\hat\dd \partial q\propto \dd q$. In particular, if
$q^{(k)}$ obeys $\hat\dd \partial q^{(k)}=-\alpha\dd q^{(k)}$,
then a
solution to the descent equations is simply
\begin{equation}
	q^{(n+k)}=\frac{\alpha!}{(n+\alpha)!}\hat\dd^nq^{(k)}  \label{eq:descent_sol}
\end{equation}
for any $n$.
The gapped condition on the family of states can be succinctly phrased as the statement that if $q^{(k)}$ is a local operator with finite correlation length, so too is $\hat{\dd}q^{(k+1)}$.
It is possible to spoil this property with a bad choice of parameter space, but for many physically reasonable choices, we expect this property to hold, and it can
in any case be checked a posteriori as we did for MPS in \cite{2024Sommer}.
Such reasonable choices likely include some suitably restricted tensor networks, a finite depth unitary preparing our state from a resource
state, sequential quantum circuits\cite{Chen_2024}, or letting the local parameter space $X_p$ be the coupling
constants on the terms in a Hamiltonian which are finitely
supported around $p$, in which case we recover the results of Ref.~\onlinecite{KS2020_higherberry}, see appendix \ref{app:KS}.

Focusing on the higher Berry curvature, if we start with the 1-form Berry connection
\begin{equation}
	F^{(1)}=\mc{A}=-\Im\braket{\psi|\dd \psi},
\end{equation}
the higher Berry flow is
\begin{equation}
	F^{(n+2)}=-\frac{1}{(n+1)!}\hat{\dd}^{n+1}\Im\braket{\psi|\dd \psi}
	\label{eq:HB_transport}
\end{equation}
and by specifying a nontrivial $n$-cochain $b$, the higher Berry curvature in a $d$ dimensional system becomes
\begin{equation}
	\Omega^{(d+2)}=\braket{F^{(d+2)},b}
	\,.
	\label{eq:HB_curvature}
\end{equation}
In $d$ dimensions, the higher Berry curvature is a flow of $(d-1)$-dimensional higher Berry curvature. See, e.g., a detailed discussion on this physical picture in Ref.\onlinecite{qpump2022}.
We remark in passing the curious fact that the formal sum of all higher Berry flows is the formal exponential of $\hat{\dd}$ acting on the connection $\sum_{n=1}^{\infty} F^{(n)}=\e^{\hat\dd} \mc{A}$.

We could equally well descend from an Abelian symmetry charge, with
the requirement that the local variations should also preserve the symmetry. For
a $\mr{U}(1)$ symmetry this leads to the Thouless pump and higher dimensional generalisations \cite{KS2020_higherthouless}.
In particular, let $\hat{Q}_p$ be the local $\mr{U}(1)$ charge operator at site $p$, from which the total charge operator is $\hat{Q}=\sum_p \hat{Q}_p$.
For any state $|\psi\rangle$ we define the charge at site $p$ as the expectation value $Q_p=\braket{\psi|\hat{Q}_p|\psi}$. The chain representing this charge distribution is the $0$-form $Q^{(0)}=\sum_p Q_p[p]$. By solving the descent equation we define the $n$-flow of charge $Q^{(n)}$, and for states with local parameter spaces, where the local parameter variations preserve the total charge, we have
\begin{equation}
	Q^{(n)}=\frac{1}{n!}\hat\dd ^n Q^{(0)}
	\,
	.\label{eq:thouless_charge_flow}
\end{equation}
On $M=\R^d$, by choosing a closed and non-exact co-chain $b$, we can define the closed $d$-form in a $d$ dimensional system as
\begin{equation}
	\tilde{Q}^{(d)}=\braket{Q^{(d)},b}
	\label{eq:d-form}
\end{equation}
which characterizes the flow of charge between boundaries at infinity for $d=1$, and the flow of the $d-1$ higher charge between boundaries at infinity for $d>1$. This is a topological invariant when integrated over a closed $d$-manifold of parameters.

A complementary perspective to these descent equations is to consider extending a quantum theory on a spacetime $M_\mr{spacetime}$ by a parameter space $X$, and figure out what higher and lower form symmetry currents\cite{gaiottoGeneralizedGlobalSymmetries2015a,kapustinHigherSymmetryGapped2017,Cordova_2020_i,Cordova_2020_ii,aloni2024spontaneously,mcnamara2020baby,tanizakiModifiedInstantonSum2020a,vandermeulen2022lowerform} mean in this context. A conserved quantity $\tilde{q}$ of the original theory is a differential $d-k$-form satisfying the conservation law $\dd_M\tilde{q}=0$, and so by an extension we seek to find a differential $d-k$ form $q$ on $M_\mr{spacetime}\times X$ satisfying $(\dd_M+\dd_X)q=0$, such that when $q$ is concentrated on $M_\mr{spacetime}$ it equals $\tilde{q}$ for that particular value of parameters. By considering the part of $q$ concentrated on the parameter space $X$, the corresponding co-homology class $[q]$ corresponds to a (higher) pump invariant of the system. In the case where there is a $k=0$-form $\mr{U}(1)$ symmetry, $\tilde{q}$ is just the ordinary $d$-form charge density. In $d=1$, the corresponding class $[q]$ on parameter space counts the amount of charge transported across a fixed point in space, as a parameter is cycled this is the Thouless pump invariant. Likewise for higher Thouless pumps. From this perspective, the higher Berry curvature is like a $k=-2$-form  $\mr{U}(1)$ symmetry current, which when concentrated to $X$ results in an invariant in degree $d+2$ of the co-homology of $X$ that is a (higher) pump of Berry curvature.
Working on the spatial lattice $\Lambda$, and considering static configurations, the differential forms on $M$ are replaced by chains, and $\dd_M$ is replaced by $-\partial$. A $k$-chain corresponds to a $d-k$-form on $M$, which explains the degree counting in the previous section. Furthermore, the connections with pumps become clear, since a $d$-chain, corresponds to a differential $0$-form on space, i.e. something that assigns values to points. For example with $\mr{U}(1)$ charge in $d=1$, $q^{(1)}$ is a $1$-chain on $\Lambda$ and a $1$-form on $X$. This corresponds to the usual current part of charge density, except we have replaced time with parameter space. Thus contracting it against a $1$-cochain corresponds to picking a spatial point $a$ and measuring just the current at this point. Analogously we obtain circulations etc. about a point for the $d>1$ case.

\subsection{Higher Berry curvatures for tensor networks}
Clearly applying the above construction to $d=1$ Matrix Product states results in exactly the higher Berry curvature explored in section \ref{sec:MPSreview} (see also Ref.\onlinecite{2024Sommer} for details), but we can generalize.
It is straightforward to apply our formulas in \eqref{eq:HB_transport} and \eqref{eq:HB_curvature} to higher dimensional tensor network states with $d\ge 2$.
The contraction of tensors which is necessary to compute the higher Berry curvature is hard to perform
due to the lack of canonical forms in higher dimensional tensor networks.
Therefore, we do not have a compact version of diagram like those for $d=1$ MPS\cite{2024Sommer}.
Nevertheless, let us give an illustration for $d=2$.
For a family of $d=2$ PEPSs, where the parmeter space at a point $p$ is just the space of tensors on this site $A_p$. The 4-form higher Berry curvature is
\begin{equation}
	\label{Eq:Omega4_a}
	\Omega^{(4)}=\langle F^{(4)},b\rangle
	=\sum_{{p\in A, q\in B,r\in C}} F_{pqr}^{(4)},
\end{equation}
where the cochain $b$ is chosen such that
we have a \emph{conical} partition of the $d=2$ lattice into three regions $A$, $B$, and $C$ (See \eqref{Eq:Omega4_c} below). The 4-form $F_{pqr}^{(4)}$ in \eqref{Eq:Omega4_a} has the explicit expression
\begin{equation}
	\label{Eq:Omega4_b}
	F^{(4)}_{pqr}=-\dd_p\,\dd_q\,\dd_r \, \rm{Im} \langle \psi|\dd \psi\rangle.
\end{equation}
Here $|\psi\rangle$ is a $d=2$ PEPS, and $\dd_p$ acts on the local tensor at site $p$ in the PEPS.

We can write down the sum of networks of transfer tensors that give the higher Berry curvature in Eqs. \eqref{Eq:Omega4_a} and \eqref{Eq:Omega4_b} graphically as follows:
\begin{widetext}
	\begin{equation}
		\label{Eq:Omega4_c}
		\Omega^{(4)}= -  \sum_{\substack{p_0\in \Lambda\\p_a\in A\\ p_b\in B\\p_c\in C}}\Im\includegraphics[valign=c]{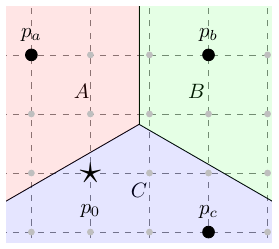},\quad
		\quad \text{where}\quad
		\begin{tikzpicture}[
				baseline ={([yshift=-4]current bounding box.center)},
				inner sep=1mm]

			\draw [thick][gray][dashed](-1.5,1.3)--(-0.5,1.3);

			\draw [thick][gray][dashed](-1.0,0.8)--(-1.0,1.8);
			\node at (-1.0,1.3) {\textcolor{lightgray}{$\bullet$}};
			\node at (-1.2,1.05) {$p$};
			\node at (-0.1,1.3) {$=$};

			\begin{scope}[yshift=5]
				\draw (1.2,2.2)--(2.2,2.2);
				\draw (0.8,1.6)--(1.8,1.6);

				\draw (0.8,1.6)--(1.2,2.2);
				\draw (1.8,1.6)--(2.2,2.2);

				\draw (0.6,1.9)--(1.0,1.9);
				\draw (2.0,1.9)--(2.4,1.9);
				\draw (1.3,1.6)--(1.1,1.3);
				\draw (1.7,2.2)--(1.9,2.5);
				\node at (1.5,1.9) {$A_p$};

				\draw (1.5,1.6)--(1.5,0.8);
			\end{scope}

			\begin{scope}[yshift=-35]
				\draw (1.2,2.2)--(2.2,2.2);
				\draw (0.8,1.6)--(1.8,1.6);

				\draw (0.8,1.6)--(1.2,2.2);
				\draw (1.8,1.6)--(2.2,2.2);

				\draw (0.6,1.9)--(1.0,1.9);
				\draw (2.0,1.9)--(2.4,1.9);
				\draw (1.3,1.6)--(1.1,1.3);
				\draw (1.7,2.2)--(1.9,2.5);
				\node at (1.5,1.9) {$\bar A_p$};
			\end{scope}

		\end{tikzpicture}
	\end{equation}
\end{widetext}
Here the unmarked gray points $\textcolor{gray}{\bullet}$  stand for the $d=2$ transfer tensor $\bb{E}^{A}_{A}$ (to compare with the 1d version in Ref.\onlinecite{2024Sommer}), the point marked $\star$ be the transfer tensor $\bb{E}^{\dd A}_{A}$, and the points marked $\bullet$ be the exterior derivative of whatever transfer tensor would the there before (with the ordering of exterior derivatives always being say the one on $p_0$ first, then $p_a$, $p_b$, and $p_c$).
Due to the lack of canonical forms in $d>1$ there is not generically an efficient way of contracting the network in \eqref{Eq:Omega4_c}. It will be interesting to apply the recently developed isometric tensor network\cite{pollman_2020}, which allows for highly efficient contraction of the tensor network, to parameterized systems in higher dimensions.

Finally, we give several remarks on the higher Berry curvatures constructed from tensor network states:

(i) It is obvious that the higher Berry curvature vanishes for a constant family of tensor network states, i.e., states that are independent of parameters in the parameter space.

(ii) The higher Berry curvature is additive under the stacking of families of states,  or more precisely the tensor product of two states, with the same underlying parameter space.

(iii) As long as the tensor network states are short-range entangled, one can argue that our higher Berry curvatures give a quantized higher Berry invariant, by following the argument in Ref.\onlinecite{KS2020_higherberry} (see their Appendix A) and using the descent equation in \eqref{eq:descent}.
The key property is that the short-range entangled states (after stacking with a suitable constant family of SRE states) can be continuously deformed to a trivial product state. Then one can relate the higher Berry invariant to the quantized Chern number of a finite system.

(iv) While the gauge independence of our higher Berry curvature for 1d canonical MPS has been studied in Ref.\onlinecite{2024Sommer},
we have not studied in detail the gauge structure of the tensor network states in higher dimensions, and it is possible that for generic tensor network states this may spoil the locality property we require of local parameter spaces, but for (semi\cite{Molnar_2018}) injective tensor networks we expect this construction to be equivalent to the construction in Ref.\onlinecite{KS2020_higherberry}, by a parent Hamiltonian construction.

\section{Higher Berry curvature calculation in $d=2$}
\label{Sec:higherD}
In $d\ge 2$ dimensions a direct calculation of the HBC becomes more involved. In our PEPS formulation this can be attributed to the more complicated contraction structure of the network. Nevertheless, in this section, we will construct a family of $d=2$ lattice models using the suspension construction\cite{qpump2022}, which we will show belongs in the higher Berry class using first the bulk-boundary correspondence, before computing the higher Berry curvature explicitly.
\subsection{Suspension construction and bulk-boundary correspondence}
\begin{figure}
	\centering
	\includegraphics{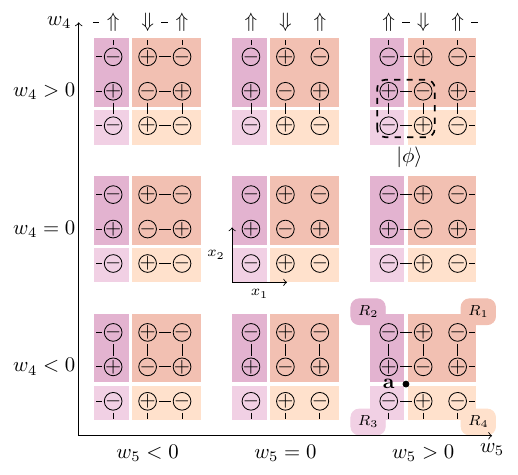}
	\caption{Illustration of $d=2$ exactly solvable model over $S^4$ with nontrivial higher Berry invariant. The $\pm$ signs indicate the sign of the local Zeeman coupling of a spin at a given site, while the lines connecting spins indicate entanglement. The regions $R_{1,2,3,4}$ are the four quadrants, and the higher Berry curvature measures the circulation around the point $\vb a$. The state $\ket{\phi}$ corresponds to the $4$-spin wave function that can be tiled to obtain the full wave function when $w_4,w_5>0$.}
	\label{fig:2d-model}
\end{figure}

The suspension construction yields a system in the nontrivial higher Berry class
in $d=2$ over $S^4$, using $d=1$ systems in the nontrivial higher Berry class
over $S^3$, in exactly the same way the Berry curvature pump in $d=1$ over $S^3$
was obtained from nontrivial $d=0$ systems over $S^2$. In particular embed
$S^4$ into $\R^5$ with coordinates $w=(\vb w,w_4,w_5)$ with $\vb w\in \R^3$, such
that $\vb w^2+w_4^2+w_5^2=1$. The lattice has sites $\vb
	x=(x_1,x_2)\in\Lambda=\Z^2$, spin $1/2$ Hilbert spaces at every $\vb x$ as
before, with coherent states along the direction $\vb n$ denoted $\ket{\vb n}$.
At the equator $w_5=0$ the system consists of Chern number pumps along the $x_2$ direction for each $x_1$. These pumps alternative in their flow, which in the concrete construction from Sec.\ref{sec:MPSreview} corresponds to alternating their local magnetic field, see figure \ref{fig:2d-model} for an illustration of the system. Away from $w_5=0$, we deform the system by coupling each chain with antiferromagnetic terms, such that for $w_5<0$ we couple the chains at $x_1\in 2\Z$ to the chains at $x_1+1$, while for $w_5>0$ the coupling is between $x_1\in 2\Z$ and $x_1-1$. In particular when $w_5=\pm 1$, the system consists entirely of spin singlet covering, with each singlet along the $x_1$ directions.
A Hamiltonian for such a system can be written
\begin{equation}
	\label{eq:H_2d}
	H_\mr{2d}(w)=\sum_{x_1\in \mathbb Z} H_{\mr{1d},x_1}((-1)^{x_1}\vb w,w_4)+H^\mr{int}_{x_1,x_1+1}(w)
\end{equation}
here $H_{\mr{1d},x_1}$ is the Hamiltonian from \eqref{eq:model_ham} taking along the $x_1\times \Z$ line, and the interaction is
$H^\mr{int}_{x_1,x_1+1}(w)=h_{\vb x}(w_5)\sum_{x_2\in\Z}\bm\sigma_{\vb x}\cdot \bm\sigma_{\vb x+(1,0)}$. We take $h_{\vb x}(w_5)=d_+(w_5)\Theta(w_5)\delta_{x_1\in 2\Z}+d_-(w_5)\Theta(-w_5)\delta_{x_1\in 2\Z+1}$, where $d_+,d_-$ vanish at $w_5=0$, and are otherwise positive. Their exact form is not important, as long as the gap never closes, which can be achieved by e.g. $d_\pm=|w_5|$.

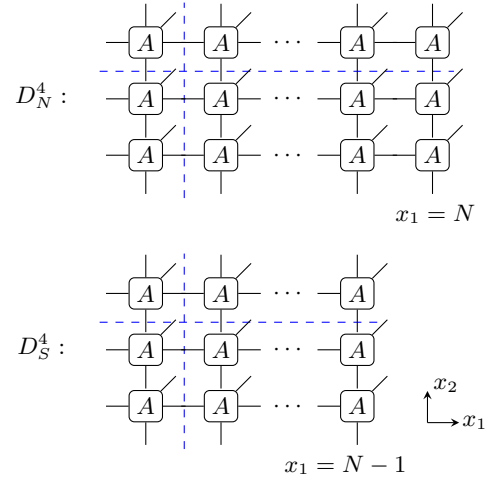
\begin{figure}[htp]
	\label{WF_2d_clutch}
	\begin{tikzpicture}[
			baseline ={([yshift=-7]current bounding box.center)},
			inner sep=1mm]
		\node at (-3.2,1.3) {$D^4_N:$};

		\draw [blue][dashed](-37pt,-2pt)--(-37pt,72pt);
		\draw [blue][dashed](-69pt,46pt)--(65pt,46pt);

		\node at (0.1,2.0) {$\cdots$};
		\node at (0.1,1.25) {$\cdots$};
		\node at (0.1,0.5) {$\cdots$};

		\node[tensor] (top) at (1.0,0.5) {$A$};
		\draw[-] (top.west)+(-0.3,0)--(top)-- (top.east)-- +(0.3,0);
		\draw[-] (top.south)+(0,-0.3)--(top);

		\node[tensor] (top) at (2.0,0.5) {$A$};
		\draw[-] (top.west)+(-0.3,0)--(top);
		\draw[-] (top.south)+(0,-0.3)--(top);

		\node[tensor] (top) at (1.0,0.5+0.75) {$A$};
		\draw[-] (top.west)+(-0.3,0)--(top)-- (top.east)-- +(0.3,0);
		\draw[-] (top.south)+(0,-0.3)--(top);

		\node[tensor] (top) at (2.0,0.5+0.75) {$A$};
		\draw[-] (top.west)+(-0.3,0)--(top);
		\draw[-] (top.south)+(0,-0.3)--(top);

		\node[tensor] (top) at (1.0,0.5+1.5) {$A$};
		\draw[-] (top.west)+(-0.3,0)--(top)-- (top.east)-- +(0.3,0);
		\draw[-] (top.south)+(0,-0.3)--(top);
		\draw[-] (top.north)+(0,0.3)--(top);

		\node[tensor] (top) at (2.0,0.5+1.5) {$A$};
		\draw[-] (top.west)+(-0.3,0)--(top);
		\draw[-] (top.south)+(0,-0.3)--(top);
		\draw[-] (top.north)+(0,0.3)--(top);

		\draw[-] (0.2+1,0.7)--(0.2+0.2+1,0.7+0.2);
		\draw[-] (0.2+2,0.7)--(0.2+0.2+2,0.7+0.2);

		\draw[-] (0.2+1,0.7+0.75)--(0.2+0.2+1,0.7+0.2+0.75);
		\draw[-] (0.2+2,0.7+0.75)--(0.2+0.2+2,0.7+0.2+0.75);

		\draw[-] (0.2+1,0.7+1.5)--(0.2+0.2+1,0.7+0.2+1.5);
		\draw[-] (0.2+2,0.7+1.5)--(0.2+0.2+2,0.7+0.2+1.5);

		\begin{scope}[xshift=-80pt]

			\node[tensor] (top) at (1.0,0.5) {$A$};
			\draw[-] (top.west)+(-0.3,0)--(top)-- (top.east)-- +(0.3,0);
			\draw[-] (top.south)+(0,-0.3)--(top);

			\node[tensor] (top) at (2.0,0.5) {$A$};
			\draw[-] (top.west)+(-0.3,0)--(top)-- (top.east)-- +(0.3,0);
			\draw[-] (top.south)+(0,-0.3)--(top);

			\node[tensor] (top) at (1.0,0.5+0.75) {$A$};
			\draw[-] (top.west)+(-0.3,0)--(top)-- (top.east)-- +(0.3,0);
			\draw[-] (top.south)+(0,-0.3)--(top);

			\node[tensor] (top) at (2.0,0.5+0.75) {$A$};
			\draw[-] (top.west)+(-0.3,0)--(top)-- (top.east)-- +(0.3,0);
			\draw[-] (top.south)+(0,-0.3)--(top);

			\node[tensor] (top) at (1.0,0.5+1.5) {$A$};
			\draw[-] (top.west)+(-0.3,0)--(top)-- (top.east)-- +(0.3,0);
			\draw[-] (top.south)+(0,-0.3)--(top);
			\draw[-] (top.north)+(0,0.3)--(top);

			\node[tensor] (top) at (2.0,0.5+1.5) {$A$};
			\draw[-] (top.west)+(-0.3,0)--(top)-- (top.east)-- +(0.3,0);
			\draw[-] (top.south)+(0,-0.3)--(top);
			\draw[-] (top.north)+(0,0.3)--(top);

			\draw[-] (0.2+1,0.7)--(0.2+0.2+1,0.7+0.2);
			\draw[-] (0.2+2,0.7)--(0.2+0.2+2,0.7+0.2);

			\draw[-] (0.2+1,0.7+0.75)--(0.2+0.2+1,0.7+0.2+0.75);
			\draw[-] (0.2+2,0.7+0.75)--(0.2+0.2+2,0.7+0.2+0.75);

			\draw[-] (0.2+1,0.7+1.5)--(0.2+0.2+1,0.7+0.2+1.5);
			\draw[-] (0.2+2,0.7+1.5)--(0.2+0.2+2,0.7+0.2+1.5);

		\end{scope}

		\node at (58pt,-8pt){$x_1=N$ };


		\begin{scope}[yshift=-95pt]
			\node at (-3.2,1.3) {$D^4_S:$};
			\draw [blue][dashed](-37pt,-2pt)--(-37pt,72pt);
			\draw [blue][dashed](-69pt,46pt)--(38pt,46pt);

			\node at (0.1,2.0) {$\cdots$};
			\node at (0.1,1.25) {$\cdots$};
			\node at (0.1,0.5) {$\cdots$};

			\node[tensor] (top) at (1.0,0.5) {$A$};
			\draw[-] (top.west)+(-0.3,0)--(top);
			\draw[-] (top.south)+(0,-0.3)--(top);

			\node[tensor] (top) at (1.0,0.5+0.75) {$A$};
			\draw[-] (top.west)+(-0.3,0)--(top);
			\draw[-] (top.south)+(0,-0.3)--(top);

			\node[tensor] (top) at (1.0,0.5+1.5) {$A$};
			\draw[-] (top.west)+(-0.3,0)--(top);
			\draw[-] (top.south)+(0,-0.3)--(top);
			\draw[-] (top.north)+(0,0.3)--(top);

			\draw[-] (0.2+1,0.7)--(0.2+0.2+1,0.7+0.2);

			\draw[-] (0.2+1,0.7+0.75)--(0.2+0.2+1,0.7+0.2+0.75);

			\draw[-] (0.2+1,0.7+1.5)--(0.2+0.2+1,0.7+0.2+1.5);

			\begin{scope}[xshift=-80pt]

				\node[tensor] (top) at (1.0,0.5) {$A$};
				\draw[-] (top.west)+(-0.3,0)--(top)-- (top.east)-- +(0.3,0);
				\draw[-] (top.south)+(0,-0.3)--(top);

				\node[tensor] (top) at (2.0,0.5) {$A$};
				\draw[-] (top.west)+(-0.3,0)--(top)-- (top.east)-- +(0.3,0);
				\draw[-] (top.south)+(0,-0.3)--(top);

				\node[tensor] (top) at (1.0,0.5+0.75) {$A$};
				\draw[-] (top.west)+(-0.3,0)--(top)-- (top.east)-- +(0.3,0);
				\draw[-] (top.south)+(0,-0.3)--(top);

				\node[tensor] (top) at (2.0,0.5+0.75) {$A$};
				\draw[-] (top.west)+(-0.3,0)--(top)-- (top.east)-- +(0.3,0);
				\draw[-] (top.south)+(0,-0.3)--(top);

				\node[tensor] (top) at (1.0,0.5+1.5) {$A$};
				\draw[-] (top.west)+(-0.3,0)--(top)-- (top.east)-- +(0.3,0);
				\draw[-] (top.south)+(0,-0.3)--(top);
				\draw[-] (top.north)+(0,0.3)--(top);

				\node[tensor] (top) at (2.0,0.5+1.5) {$A$};
				\draw[-] (top.west)+(-0.3,0)--(top)-- (top.east)-- +(0.3,0);
				\draw[-] (top.south)+(0,-0.3)--(top);
				\draw[-] (top.north)+(0,0.3)--(top);

				\draw[-] (0.2+1,0.7)--(0.2+0.2+1,0.7+0.2);
				\draw[-] (0.2+2,0.7)--(0.2+0.2+2,0.7+0.2);

				\draw[-] (0.2+1,0.7+0.75)--(0.2+0.2+1,0.7+0.2+0.75);
				\draw[-] (0.2+2,0.7+0.75)--(0.2+0.2+2,0.7+0.2+0.75);

				\draw[-] (0.2+1,0.7+1.5)--(0.2+0.2+1,0.7+0.2+1.5);
				\draw[-] (0.2+2,0.7+1.5)--(0.2+0.2+2,0.7+0.2+1.5);

			\end{scope}

			\begin{scope}[xshift=55pt,yshift=0pt]
				\draw [>=stealth,->] (0pt,8pt)--(12pt,8pt);
				\draw [>=stealth,->] (0pt,8pt)--(0pt,20pt);

				\node at (18pt,8pt){$x_1$ };
				\node at (7pt,22pt){$x_2$ };

			\end{scope}

			\small
			\node at (24pt,-8pt){$x_1=N-1$ };
			\normalsize

		\end{scope}

	\end{tikzpicture}
	\caption{Families of $d=2$ PEPSs corresponding to the ground states of $d=2$ lattice models in \eqref{eq:H_2d} with a spatial boundary terminated at layer $x_1=N$ (top) and $x_1=N-1$ (bottom), where $N\in 2\mathbb Z$. The blue dashed lines indicate the steps in the functions $f_1(p)=\Theta(x_1(p)-a_1)$ and $f_2(p)=\Theta(x_2(p)-a_2)$.
	}
\end{figure}

The $d=2$ higher Berry curvature is the pump of a pump, and so is equal to the circulation of Berry curvature current $F^{(4)}$ evaluated around plaquettes enclosing a ``core'' at  $\vb a\in \Lambda+(1/2,1/2)$. To properly capture the circulation around $\vb a$ one can chose many equivalent expressions for $b_{\vb a}$, but for computational purposes we find it most convenient to compose it of the properly antisymmetrized $d=1$ boundaries, as opposed to the conical choice mentioned before: let $f_i^p=\Theta(x_i(p)-a_i)$ be a step function and define
\begin{align}
	b_{\vb a} & =(f_1^{p_1}-f_1^{p_0})(f_2^{p_2}-f_2^{p_1})(p_0p_1p_2)                                                 \\
	          & =\delta\left[\frac 12(f_1^{p_0}+f_1^{p_1})(f_2^{p_1}-f_2^{p_0})(p_0p_1)\right]=\delta \tilde b_{\vb a}
\end{align}
where $\tilde b_{\vb a}$ is a $1$-cochain that is not cocontrolled in a system without a boundary.
Take regions $R_i$ to be the $i$'th quadrant centered at $\vb a$, then $b_{\vb a}^{p_0p_1p_2}$ vanishes unless all three point are in different regions. With this cochain, the higher Berry curvature is
\begin{equation}
	\Omega^{(3)}=\braket{F^{(4)},b_{\vb a}}
\end{equation}
where $F^{(n)}$ is defined by \eqref{eq:HB_transport} for arbitrary $n$. We can immediately see that our family has to be in the higher Berry class, by imposing a spatial boundary on the right side of the $d=2$ system.
That is, the system is defined by retaining only those lattice sites with $x_1\leq N$.
In our lattice model in \eqref{eq:H_2d}, one can find that $F^{(4)}_{pqr}$ is only nonzero when $p$, $q$, $r$ are near the point $\vb a$. In a general gapped ground state, we expect the locality of $F^{(4)}$ is still true. Due to this locality of $F^{(4)}$ in our lattice model, the higher Berry curvature $\Omega^{(4)}$ is insensitive to the existence of a boundary if we choose $\vb a$ deep in the bulk. Using the existence of boundary, one can find
\begin{equation}
	\label{Omega4omega3}
	\Omega^{(4)}=\dd \braket{F^{(3)},\tilde b_\vb a}=\dd \omega^{(3)}.
\end{equation}
Noting that $\tilde b_{\vb a}^{p_0p_1}$ is non-zero whenever $x_2(p_0)$ and $x_2(p_1)$ lie on opposite sides of $a_2$, and if $x_1(p_0) > a_1$ or $x_1(p_1) > a_1$, we see that $\omega^{(3)}$ is the 3-form boundary higher Berry curvature, where the boundary
consists of all lattice sites with $x_1>a_1$
\cite{qpump2022}. When the boundary is decoupled from the bulk, $\omega^{(3)}$ is the closed $3$-form we evaluated in section \ref{sec:MPSreview}. However, $\omega^{(3)}$ is not globally well defined, because the boundary will become gapless for certain parameters. To evaluate $\Omega^{(4)}$, we therefore cover $S^4$ with two charts $D_{N/S}^4=\set{w\in S^4,\pm w_5\geq 0}$, with common boundary of opposite orientation $S^3=\pm\partial D^{4}$.
On the two charts $D_N^4$ and $D_S^4$, we impose different boundaries, and consider the two families of PEPS that correspond to the
ground states of $H_{\mr{2d}}$ in \eqref{eq:H_2d} with a boundary imposed at $x_1=N$ and $x_1=N-1$ (where $N\in 2\mathbb Z$) respectively, as seen in Fig.\ref{WF_2d_clutch}. With this choice, the ground states are always gapped, and $\omega^{(3)}_{N/S}$ in \eqref{Omega4omega3} are well defined on both $D_N^4$ and $D_S^4$.
So we may evaluate
\begin{equation}
	\int_{S^4}\Omega^{(4)}=\int_{S^3}\omega^{(3)}_N-\omega_S^{(3)}.\label{eq:clutching}
\end{equation}

Since the $d=2$ system becomes decoupled $d=1$ systems at $w_5=0$, the PEPS becomes decoupled $d=1$ MPS. One can find that $\omega_N^{(3)}-\omega_S^{(3)}$
in \eqref{eq:clutching} equals the closed 3-form higher Berry curvature $\Omega^{(3)}$ for the $d=1$ MPS along $x_1=N$, and so the higher Berry class is as promised
\begin{equation}
	\int_{S^4} \Omega^{(4)}=\int_{S^3} \Omega^{(3)}=2\pi.
\end{equation}

One can repeat this procedure to even higher dimensions. For the lattice models obtained from suspension construction in Ref.~\onlinecite{qpump2022}, one can show that our higher Berry curvatures constructed from tensor networks give the same higher Berry invariants as those from Kapustin and Spodyneiko's construction\cite{KS2020_higherberry}.

\subsection{Bulk calculation}
In addition to the bulk-boundary correspondence, it is also possible to calculate the higher Berry curvature directly in terms of the wave function, by using a PEPS parameterization. In particular the unit cell is $2\times 2$ sites, and we use tensors $A$, which act on a virtual space with basis $|\cdot)),|1)),|2)),|3)),|4))$. Since there can only be higher Berry curvature when there is entanglement between three of the regions $R_i$, we may focus on the parameters $w_4,w_5>0$. Let $i,j$ denote $1,2,3,4$, and we can write the $2\times 2$ unit cell of PEPS tensors in this parameter range in terms of MPS tensors
\begin{equation*}
	\includegraphics{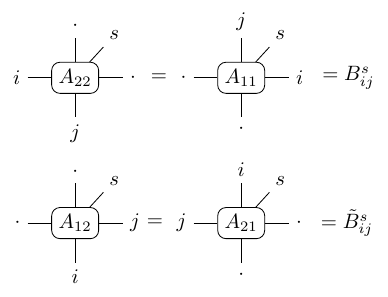}
\end{equation*}
Entanglement occurs only within this unit cell, and so the total wave function is a product of plaquette states $\ket{\phi}$, which are related to the MPS tensors as $\braket{s_{11},s_{21},s_{22},s_{12}|\phi}=\tr B^{s_{11}}\tilde{B}^{s_{21}}B^{s_{22}}\tilde{B}^{s_{12}}$.
While one could in principle solve the ground state for an particular interpolating Hamiltonian, since the coupling between pumps is arbitrary, we might as well just write down a wave function over $S^4$ which couples the pumps correctly.
We take the MPS to have the form of the vertical pump stacked with a horizontal pump, i.e.
\begin{align}
	B^{s}         & =\begin{pmatrix}
		                 \braket{s|\vb w} & \braket{s|-\vb w} & 0                 \\
		                 0                & 0                 & 0                 \\
		                 0                & 0                 & \braket{s|\vb w}  \\
		                 0                & 0                 & \braket{s|-\vb w}
	                 \end{pmatrix}                                        \\
	\tilde{B}^{s} & =\begin{pmatrix}
		                 \alpha\braket{s|-\vb w} & 0 & 0                               & 0                               \\
		                 -\beta\braket{s|\vb w}  & 0 & 0                               & 0                               \\
		                 0                       & 0 & \i^{1/2}\delta\braket{s|-\vb w} & -\i^{1/2}\gamma\braket{s|\vb w}
	                 \end{pmatrix}
\end{align}
where $\alpha,\beta,\gamma,\delta\in\R^{\geq 0}$, and normalization requires $1=(\alpha^2+\beta^2)^2+(\delta^2+\gamma^2)^2$.
Let $\omega^{(2)}$ be half the solid angle area element of the $S^2$ corresponding to the directions $\vb w$.
Using equation \eqref{Eq:Omega4_c}, and replacing the tripartition with the cochain $b_{\vb a}$, the higher Berry curvature is
\begin{align}
	\label{eq:HBC_computation_2d}
	\Omega^{(4)} & =\Theta(w_4)\Theta(w_5)\Im \tr \bb{E}^{B}_{B}\dd\bb{E}^{\tilde{B}}_{\tilde{B}}\bb{E}^{\dd B}_{\dd B}\dd \bb{E}^{\tilde B}_{\tilde B} \\
	             & =2\Theta(w_4)\Theta(w_5)\left[\dd\alpha^2\dd\beta^2+\dd\gamma^2\dd\delta^2\right]\omega^{(2)}.
\end{align}
As mentioned before there are some arbitrariness present in the choice of $\alpha,\beta,\gamma,\delta$, relating to the choice of interaction for which this is the ground state. However, no matter the choice, the higher Berry invariant is always $2\pi$ quantized, as can be checked by Stokes theorem. In particular by the suspension construction, on the boundary $\Gamma_1: w_4=0,w_5\in[0,1]$ $\alpha=\beta=0$, while on $\Gamma_2: \,w_5=0,w_4\in[0,1]$ it holds that $\alpha^2=1-\beta^2$ and $\alpha$ goes from $1$ to $1/\sqrt{2}$, and finally on $\Gamma_3:\vb w=0$, we have $\alpha^2=\beta^2$ and $\alpha$ goes from $1/\sqrt{2}$ to $0$. Similarly for $\gamma,\delta$ with $\Gamma_{1,2}$ swapped. Thus the integral becomes
\begin{align*}
	\int\Omega^{(4)} & =4\pi\left[\int_{w_{4,5}>0}\dd\alpha^2\dd\beta^2+\dd\gamma^2\dd\delta^2\right]            \\
	                 & =4\pi\left[\int_{\Gamma_1+\Gamma_2+\Gamma_3}\alpha^2\dd\beta^2+\gamma^2\dd\delta^2\right] \\
	                 & =4\pi\left[0+\frac{3}{8}-\frac{1}{8}+\frac{3}{8}+0-\frac{1}{8}\right]                     \\
	                 & =2\pi.
\end{align*}
To illustrate, we find the simplest choice is obtained by defining $\hat{w}_5=w_5/\sqrt{w_4^2+w^2_5}$, $\hat{w}_4=w_4/\sqrt{w_4^2+w^2_5}$, and letting
\begin{align*}
	\alpha=\sqrt{\frac{1}{2}\hat{w}_4\left(1+|\vb w|\right)}; & \qquad \beta =\sqrt{\frac{1}{2}\hat{w}_4\left(1-|\vb w|\right)} \\
	\delta=\sqrt{\frac{1}{2}\hat{w}_5\left(1+|\vb w|\right)}; & \qquad \gamma=\sqrt{\frac{1}{2}\hat{w}_5\left(1-|\vb w|\right)}
\end{align*}
hence
\begin{equation}
	\Omega^{(4)}=\Theta(w_4)\Theta(w_5)\left(-\hat{w}_4\dd \hat{w}_4+\hat{w}_5\dd\hat{w}_5\right)\dd |\vb w|\,\omega^{(2)},
\end{equation}
which can be easily computed directly, and as promised $\int_{S^4}\Omega^{(4)}=\int_{S^2}\omega^{(2)}=2\pi$.

\section{Discussion}
\label{sec:Discuss}
In this work, we propose a systematic wavefunction based approach to construct higher Berry curvatures and higher Thouless pumps for a parametrized family of systems with local parameter spaces. We apply this approach to parametrized families of tensor networks
that correspond to the ground states of short-range entangled systems. We construct $(d+2)$-form higher Berry curvatures for a $d$ dimensional system, as well as $d$-form that characterizes the higher Thouless pumps in a $d$ dimensional system with $\mr{U}(1)$ symmetry. 
Our formulas are suitable for both analytical studies of exactly solvable models and numerical studies of general lattice models, and we illustrate these applications with several examples.

\medskip
In the following, we mention several interesting future problems.
One future direction is to apply our approach to the study of parametrized topologically ordered systems. 
In this case, it is expected that the higher Berry invariants 
as well as the higher Thouless pump may be fractional. See, e.g., a recent study on this fractional phenomenon in parametrized families of systems from the field theory point of view\cite{Hsin_2023}.

It is also desirable to give a rigorous proof of the quantization of higher Berry curvatures in our construction.
For $d=1$ MPS, the proof has been given for the translationally invariant MPS in Ref.\onlinecite{2024Sommer}, where it was shown that our higher Berry curvature corresponds to the curvature for the gerbe in translationally invariant MPSs. In higher dimensions, one can argue that our higher Berry curvatures for short-range entangled states also give a quantized higher Berry invariant by following the same argument in Ref.\onlinecite{KS2020_higherberry}, but a rigorous proof is still needed.

Another future direction is to study how to detect higher Berry curvatures in experiments.
 Even for $d=1$ parametrized systems, the expressions of higher Berry curvatures constructed in this work are different from those obtained from the Hamiltonians \cite{KS2020_higherberry} and from the numerical approach\cite{2023Shiozaki}. 
It seems to us there is not a canonical way of defining higher Berry curvatures, in contrast to the canonical 2-form Berry curvature, knowing only the family of states. In particular, they correspond to different ways of embedding $X$ into some parameter space with locality $\prod_p X_p$. 
Physically, different ways of constructing higher Berry curvatures may correspond to different protocols to detect them. It is an interesting future work to study how to detect such higher Berry curvatures in experiments. 

\bigskip
\textit{Note added}: While completing this manuscript, we became aware of an upcoming related work Ref.\onlinecite{2024Ryu} to appear on arXiv on the same day.

\acknowledgments

We thank Michael Hermele and Cristian Batista for helpful discussions and general comments and Daniel Parker for insights into MPS.
This work was supported in part by the Simons Collaboration on Ultra-Quantum Matter, which is a grant from the Simons Foundation (618615, XW, AV; 651440, XW, MH).
\appendix

\section{Higher Berry curvatures for locally parametrized states beyond tensor networks}
\label{Sec:Unitaries}

Our approach of constructing higher Berry curvatures and higher Thouless pumps not only works for tensor network states as illustrated in the main text, but also works for other types of locally parametrized states. In this appendix, we give an example on the family of states that are generated by applying parametrized local unitary operators on a reference state $|\psi_0\rangle$:
\be
\label{Psi_U}
|\psi(\{\lambda_p\})\rangle=U(\{\lambda_p\})|\psi_0\rangle,
\ee
where $U(\{\lambda_p\})$ denotes the unitaries that depend on local parameters in the local parameter spaces $X_p$ for arbitrary lattice sites $p\in \Lambda$. 
Hereafter, for simplicity of writing we will write \eqref{Psi_U} as $|\psi\rangle=U|\psi_0\rangle$.
Then the total variation of the family of states decomposes into a sum over local variations
\be
\dd |\psi\rangle=\sum_p \dd_p|\psi\rangle=\sum_p (\dd_p U)|\psi_0\rangle.
\ee
Note that in general one cannot choose unitary operators $U$ that are smooth everywhere in the whole parameter space. In this case, we can cover the parameter space by several charts. Then the unitary operators can be smooth everywhere on each chart. 

Since the unitary operators $U$ in \eqref{Psi_U} depend on both the lattice sites and the external parameters, this allows us to define the simplex chain and differential forms following the procedures in Sec.\ref{Sec:GeneralConstruct}. More explicitly, to construct the higher Berry curvatures, one can start from the regular 1-form $\mathcal A$ as
\be
\label{eq:1-form_App}
\mathcal A=-\Im\langle \psi|\dd\psi\rangle=-\Im\sum_p\langle \psi_0 | (\dd_p U)|\psi_0\rangle.
\ee
Then the $n$-chain valued $(n+2)$-form $F^{(n+2)}$ can be obtained from \eqref{eq:HB_transport}, based on which one can construct the $(n+2)$-form higher Berry curvature according to \eqref{eq:HB_curvature}.

In the following, let us illustrate this construction with the exactly solvable model in Sec.\ref{sec:MPSreview}. We consider the same configuration in \eqref{eq:H_config}. 
Let us focus on $w_4\ge 0$ first. For each dimer on sites 
$p\in 2\mathbb Z-1$ and $(p+1)\in 2\mathbb Z$, the wavefunction can be expressed as
\be
\label{eq:dimer_wf}
|\psi\rangle_{p,p+1}=
\Big[U_p(\theta,\phi)\otimes U_{p+1}(\theta,\phi)\Big] U_{p,p+1}(\alpha)\, |\psi_0\rangle.
\ee
The total wavefunction is a tensor product of wavefunctions on each dimer.
Here $|\psi_0\rangle=|\uparrow\downarrow\rangle$ is the ground state of the dimer at $w_4=0$ and $\vb{w}=(w_1,w_2,w_3)=(0,0,1)$.
The unitary operator $U_p$ acts on the local Hilbert space $\mathcal H_p$, and the parameters in $U_p$ belong to the local parameter space $X_p$. More explicitly, we have
\be
\label{eq:Utheta}
U_p=U_{p+1}=U(\theta,\phi)=\begin{pmatrix}
\cos\frac{\theta}{2} &-\sin\frac{\theta}{2}e^{-i\phi}\\
\sin\frac{\theta}{2}e^{i\phi} &\cos\frac{\theta}{2}
\end{pmatrix},
\ee
where $0\le \theta\le \pi, \, 0\le \phi\le2\pi$. 
Here $U(\theta,\phi)$ rotates the eigenstates of $\sigma^z$ to eigenstates of $\vb{w}\cdot \bm\sigma$.
The unitary operator $U_{p,p+1}$, which acts on the Hilbert space $\mathcal H_p\otimes \mathcal H_{p+1}$, has the expression
\be
\begin{split}
&U_{p,p+1}(\alpha)=
|\uparrow\uparrow\rangle\langle \uparrow\uparrow|
+
|\downarrow\downarrow\rangle\langle \downarrow\downarrow|\\
&+\sqrt{\frac{1+\sin\alpha}{2}}
(|\uparrow\downarrow\rangle\langle \uparrow\downarrow |-|\downarrow\uparrow\rangle\langle \downarrow\uparrow |)\\
&-\sqrt{\frac{1-\sin\alpha}{2}}
(|\uparrow\downarrow\rangle\langle \downarrow \uparrow |+|\downarrow\uparrow\rangle\langle \uparrow \downarrow |).
\end{split}
\ee
where $0\le \alpha\le \pi/2$.
Now we have the freedom to decide $\alpha$ in $U_{p,p+1}(\alpha)$ belongs to the parameter spaces $X_p$ or $X_{p+1}$. In fact, the same `gauge freedom' also appears in the MPS formalism (See Appendix.\ref{Sec:ThoulessPump}).
Choosing $\alpha\in X_p$ here corresponds to the right canonical MPS, and choosing $\alpha\in X_{p+1}$ corresponds to the left canonical. For either choice, we can evaluate the higher Berry curvature  $\Omega^{(3)}$ by using \eqref{eq:F3} and \eqref{Omega3}, where the 1-form $\mathcal A$ is now given in \eqref{eq:1-form_App}. 
We find that for either $\alpha\in X_p$ or $\alpha\in X_{p+1}$, the higher Berry curvature $\Omega^{(3)}$ has the same expression as the MPS result in \eqref{eq:3-form_toy}.

Next, for $w_4< 0$, since the spins at sites $p\in 2\mathbb Z-1$ and $(p+1)\in 2\mathbb Z$ are decoupled from each other, there is no Berry curvature flow across the cut (dashed line in \eqref{eq:H_config}). Repeating the above procedure, one can find the higher Berry curvature vanishes for $w_4< 0$, which again agrees with the MPS result. 

As a remark, for the purpose of illustrating our construction, 
we choose a very simple family of unitary operators in the above discussion. These unitary operators are not smooth everywhere over the parameter space $S^3$. One can find they are not well defined at $\alpha=0$ and $\theta=\pi$. More rigorously, one should cover $S^3$ with open covers. On each open cover, the unitary operators can be smooth and well defined everywhere. For the toy model we consider here, one should be able to write down the smooth unitary operators on each open cover by following a similar procedure in Ref.\onlinecite{2024Sommer}.

\section{Relation to KS formulas}
\label{app:KS}
In Ref.~\onlinecite{KS2020_higherberry}, Kapustin and Spodyneiko first introduced Kitaev's idea of the
higher Berry curvature to the literature. They formalise the notion using coarse
geometry as we do here, and consider gapped Hamiltonians whose ground states are short range entangled.
While the choice of local parent Hamiltonian is immaterial to the invariant, which is a characteristic only of the short-range entangled
states, their formulation requires picking a parent Hamiltonian. In this appendix we show that their
formulation can derived in the language of local parameter spaces.

Consider as before a lattice $\Lambda$, and a Hilbert space $\mc{H}$ with a tensor
product factorisation over $\Lambda$, so that $\mc{H}=\otimes_{p\in \Lambda} \mc{H}_p$. 
For each $p\in \Lambda$, consider the space of finitely ranged self-adjoint operators $X_p$ with maximal range $r$.
That is, if $H_p,H_q$ are such operators, and $d(p,q)>2r$, then they must
commute $[H_p,H_q]=0$. We can define a finitely ranged Hamiltonian as a sum of such operators $H=\sum_p
H_p$. Let the total parameter space $X$ be the subset of $\prod_{p\in
\Lambda}X_p$ such that the ground state is unique and gapped. Then the
Hamiltonian implicitly defines a map $X\to\mc{H}$ taking any set of parameters
to the gapped ground state. The Berry curvature of such a ground
state is extensive and hence divergent, but by using a complex contour integral encircling
the ground state energy it formally takes the form
\begin{align}
        F=\frac{\i}{2}\int\frac{\dd z}{2\pi \i}\tr[G\dd HG^2 {\dd}
    H],
\end{align}
where $\dd H$ is the exterior derivative of the Hamiltonian, and $G(z)=(z-H)^{-1}$
is the resolvant. Since it is located over all of space we can view it as a
$-1$-chain, and find a local Berry curvature $0$-chain, which has $F$ as a
boundary. In particular using the simplex-exterior derivative their choice can be written\cite{KS2020_higherberry}
\begin{equation}
    F^{(2)}=\frac{\i}{2}\int \frac{\dd z}{2\pi\i}\tr\left[G\dd H G^2\hat\dd    H\right].
\end{equation}
We can check that $\hat\dd\partial F^{(2)}=-2\dd F^{(2)}$, so by equation
\eqref{eq:descent_sol} the higher Berry flow is simply 
\begin{equation}
    F^{(n+2)}=\frac{2}{(n+2)!}\hat\dd^nF^{(2)}.
\end{equation}
Using $\hat\dd G=G\hat\dd H G$, we find
\begin{equation}
        \hat\dd^nF^{(2)}=\i\sum_{l=0}^n c^n_l \int\frac{\dd z}{2\pi\i}\tr\left[ \dd H
        (G\hat\dd H)^l G^2 (\hat\dd H G)^{n-l+1}\right]
\end{equation}
where $c^n_l=(-)^nn! (n-l+1)/2$, where by the product of two simplexes $[p_0\cdots p_n][q_0\cdots q_m]$ is meant the simplex $[p_0\cdots p_nq_0\cdots q_m]$. This is of course not generically a well defined operation on chains with finite correlation length, but it is on the free chains without that constraint. We still use the product since by the assumption on local parameter spaces, the total expression is controlled.
To prove the formula, we just need the simplex-exterior derivative of any given term
\begin{align*}
    \hat\dd \tr &\left[ \dd H
    (G\hat\dd H)^a G^2 (\hat\dd HG)^{b}\right]\\
    =&-(a+1)\tr\left[ \dd H
(G\hat{\dd}H)^{a+1}G^2(\hat\dd H G)^{b}\right]\\
                                               &-(b+1) \tr \left[\dd H
                                               (G\hat{\dd}H)^{a}G^2(\hat\dd H
                                           G)^{b+1}\right].
    \end{align*}
Finally the higher Berry flow becomes 
\begin{align}
    F^{(n+2)}=&\frac{(-)^n\i}{(n+2)(n+1)}\sum_{l=0}^n(n-l+1)\nonumber \\ 
    &\int \frac{\dd
    z}{2\pi\i}\tr\left[ \dd H
        (G\hat\dd H)^l G^2 (\hat\dd H G)^{n-l+1}\right]. \label{eq:KS}
\end{align}
Expanding the simplex indices, this is the higher Berry curvature result in Ref.~\onlinecite{KS2020_higherberry}.

\section{Higher Thouless charge pump in exactly solvable models}
\label{Sec:ThoulessPump}

In this appendix, we apply our formulas in Eqs.\eqref{eq:thouless_charge_flow} and \eqref{eq:d-form} to lattice systems with a global $U(1)$ symmetry. These parametretrized systems include  the well known Thouless charge pump in $1d$ systems, and higher dimensional systems with higher Thouless charge pump. The topological invariants in terms of a family of Hamiltonians in higher Thouless pumps were recently studied in Ref.~\onlinecite{KS2020_higherthouless}.

\subsection{Thouless charge pump in $d=1$}

We consider a $d=1$ lattice models with a global $\mr{U}(1)$ symmetry, with the parameter space $X=S^1$.
This model has the interesting feature of $\mr{U}(1)$ charge Thouless pump as we adiabatically change the parameters along $X=S^1$.

Let $w=(w_1,w_2)\in S^1\subseteq \R^2$ be the standard embedding of the unit circle, and $\alpha\in\R/2\pi\Z$ be the corresponding angle, which is a parameterization via the map $\alpha\mapsto (\cos\alpha,\sin\alpha)$.
The family of Hamiltonians we consider are very similar to the those in Sec.\ref{sec:MPSreview}. In fact it may be constructed by restricting the Berry curvature pump to the $\theta=0,\pi$ part of the parameter space, after a slight rearrangement of parameters. As such it has the form 
\begin{equation}
\label{H_1d_U1}
H_\mr{1d}(w)=\sum_{p\in\mathbb Z} H_p^\mr{onsite}(w)+H^{\mr{int}}_{p,p+1}(w),
\end{equation}
where the terms are now given by
\begin{equation}
\label{H_1body_U1}
H_p^\mr{onsite}(w)=  (-1)^p \, w_1\sigma_p^3,\qquad H^{\mr{int}}_{p,p+1}=g_p(w_2)\bm{\sigma}_p\cdot\bm{\sigma}_{p+1}
\end{equation}
and $g_p$ is defined by equation \eqref{eq:gp}.
Pictorially, the Hamiltonians
for different values of $\alpha\in [0,2\pi]$ can be visualized as follows:  
 \begin{equation}
\label{H_configU1}
\small
\begin{tikzpicture}

\node at (-59pt,0pt){$\alpha=\pi/4$:};


\notsotiny
\draw (-20pt,0pt) circle (4.5pt);
\node at (-20pt,0pt){$+$};
\draw [thick](4.5-35pt,0pt)--(15.5-40pt,0pt);

\draw (0pt,0pt) circle (4.5pt);
\node at (0pt,0pt){$-$};
\draw [thick](4.5pt,0pt)--(15.5pt,0pt);
\draw (20pt,0pt) circle (4.5pt);
\node at (20pt,0pt){$+$};

\draw (40pt,0pt) circle (4.5pt);
\node at (40pt,0pt){$-$};
\draw [thick](44.5pt,0pt)--(55.5pt,0pt);
\draw (60pt,0pt) circle (4.5pt);
\node at (60pt,0pt){$+$};

\draw (80pt,0pt) circle (4.5pt);
\node at (80pt,0pt){$-$};
\draw [thick](84.5pt,0pt)--(95.5pt,0pt);
\draw (100pt,0pt) circle (4.5pt);
\node at (100pt,0pt){$+$};

\draw (120pt,0pt) circle (4.5pt);
\node at (120pt,0pt){$-$};
\draw [thick](124.5pt,0pt)--(135.5pt,0pt);
\draw (140pt,0pt) circle (4.5pt);
\node at (140pt,0pt){$+$};

\begin{scope}[yshift=44pt]
\small
\node at (-61pt,0pt){$\alpha=3\pi/4$:};

\draw [dashed](-10+20pt,120pt)--(-10+20pt,-70pt);

\notsotiny
\draw (-20pt,0pt) circle (4.5pt);
\node at (-20pt,0pt){$-$};
\draw [thick](4.5-35pt,0pt)--(15.5-40pt,0pt);

\draw (0pt,0pt) circle (4.5pt);
\node at (0pt,0pt){$+$};
\draw [thick](4.5pt,0pt)--(15.5pt,0pt);
\draw (20pt,0pt) circle (4.5pt);
\node at (20pt,0pt){$-$};

\draw (40pt,0pt) circle (4.5pt);
\node at (40pt,0pt){$+$};
\draw [thick](44.5pt,0pt)--(55.5pt,0pt);
\draw (60pt,0pt) circle (4.5pt);
\node at (60pt,0pt){$-$};

\draw (80pt,0pt) circle (4.5pt);
\node at (80pt,0pt){$+$};
\draw [thick](84.5pt,0pt)--(95.5pt,0pt);
\draw (100pt,0pt) circle (4.5pt);
\node at (100pt,0pt){$-$};

\draw (120pt,0pt) circle (4.5pt);
\node at (120pt,0pt){$+$};
\draw [thick](124.5pt,0pt)--(135.5pt,0pt);
\draw (140pt,0pt) circle (4.5pt);
\node at (140pt,0pt){$-$};

\end{scope}

\begin{scope}[yshift=-22pt]
\small
\node at (-53pt,0pt){$\alpha=0$:};

\notsotiny
\draw (-20pt,0pt) circle (4.5pt);
\node at (-20pt,0pt){$+$};

\draw (0pt,0pt) circle (4.5pt);
\node at (0pt,0pt){$-$};

\draw (20pt,0pt) circle (4.5pt);
\node at (20pt,0pt){$+$};

\draw (40pt,0pt) circle (4.5pt);
\node at (40pt,0pt){$-$};

\draw (60pt,0pt) circle (4.5pt);
\node at (60pt,0pt){$+$};

\draw (80pt,0pt) circle (4.5pt);
\node at (80pt,0pt){$-$};

\draw (100pt,0pt) circle (4.5pt);
\node at (100pt,0pt){$+$};

\draw (120pt,0pt) circle (4.5pt);
\node at (120pt,0pt){$-$};

\draw (140pt,0pt) circle (4.5pt);
\node at (140pt,0pt){$+$};

\end{scope}

  \begin{scope}[yshift=132pt]
  
  \small

\node at (-61pt,0pt){$\alpha=7\pi/4$:};

\notsotiny

\draw (-20pt,0pt) circle (4.5pt);
\node at (-20pt,0pt){$+$};

\draw (0pt,0pt) circle (4.5pt);
\node at (0pt,0pt){$-$};
\draw [thick](4.5-20pt,0pt)--(15.5-20pt,0pt);
\draw (20pt,0pt) circle (4.5pt);
\node at (20pt,0pt){$+$};

\draw (40pt,0pt) circle (4.5pt);
\node at (40pt,0pt){$-$};
\draw [thick](24.5pt,0pt)--(35.5pt,0pt);
\draw (60pt,0pt) circle (4.5pt);
\node at (60pt,0pt){$+$};

\draw (80pt,0pt) circle (4.5pt);
\node at (80pt,0pt){$-$};
\draw [thick](64.5pt,0pt)--(75.5pt,0pt);
\draw (100pt,0pt) circle (4.5pt);
\node at (100pt,0pt){$+$};

\draw (120pt,0pt) circle (4.5pt);
\node at (120pt,0pt){$-$};
\draw [thick](104.5pt,0pt)--(115.5pt,0pt);
\draw (140pt,0pt) circle (4.5pt);
\node at (140pt,0pt){$+$};
\draw [thick](144.5pt,0pt)--(150.5pt,0pt);
   \end{scope}
   
     \begin{scope}[yshift=+110pt]

\draw [thick](-20pt,0pt)--(0pt,0pt);
\draw [thick](20pt,0pt)--(40pt,0pt);
\draw [thick](60pt,0pt)--(80pt,0pt);
\draw [thick](100pt,0pt)--(120pt,0pt);
\draw [thick](140pt,0pt)--(150pt,0pt);

\node at (140pt,0pt){$\bullet$};
\node at (120pt,0pt){$\bullet$};
\node at (100pt,0pt){$\bullet$};
\node at (80pt,0pt){$\bullet$};
\node at (60pt,0pt){$\bullet$};
\node at (40pt,0pt){$\bullet$};
\node at (20pt,0pt){$\bullet$};
\node at (0pt,0pt){$\bullet$};
\node at (-20pt,0pt){$\bullet$};

\small
\node at (-60pt,0pt){$\alpha=3\pi/2$:};
\end{scope}

     \begin{scope}[yshift=22pt]
     
\draw [thick](-30pt,0pt)--(-20pt,0pt);
\draw [thick](0pt,0pt)--(20pt,0pt);
\draw [thick](40pt,0pt)--(60pt,0pt);
\draw [thick](80pt,0pt)--(100pt,0pt);
\draw [thick](120pt,0pt)--(140pt,0pt);

\node at (140pt,0pt){$\bullet$};
\node at (120pt,0pt){$\bullet$};
\node at (100pt,0pt){$\bullet$};
\node at (80pt,0pt){$\bullet$};
\node at (60pt,0pt){$\bullet$};
\node at (40pt,0pt){$\bullet$};
\node at (20pt,0pt){$\bullet$};
\node at (0pt,0pt){$\bullet$};
\node at (-20pt,0pt){$\bullet$};

\small
\node at (-58pt,0pt){$\alpha=\pi/2$:};

\begin{scope}[yshift=-85pt]

\end{scope}

\end{scope}


\begin{scope}[yshift=66pt]
\small
\node at (-54pt,0pt){$\alpha=\pi$:};

\notsotiny
\draw (-20pt,0pt) circle (4.5pt);
\node at (-20pt,0pt){$-$};

\draw (0pt,0pt) circle (4.5pt);
\node at (0pt,0pt){$+$};

\draw (20pt,0pt) circle (4.5pt);
\node at (20pt,0pt){$-$};

\draw (40pt,0pt) circle (4.5pt);
\node at (40pt,0pt){$+$};

\draw (60pt,0pt) circle (4.5pt);
\node at (60pt,0pt){$-$};

\draw (80pt,0pt) circle (4.5pt);
\node at (80pt,0pt){$+$};

\draw (100pt,0pt) circle (4.5pt);
\node at (100pt,0pt){$-$};

\draw (120pt,0pt) circle (4.5pt);
\node at (120pt,0pt){$+$};

\draw (140pt,0pt) circle (4.5pt);
\node at (140pt,0pt){$-$};

\end{scope}

 \begin{scope}[yshift=88pt]
    \small
\node at (-61pt,0pt){$\alpha=5\pi/4$:};

\notsotiny

\draw (-20pt,0pt) circle (4.5pt);
\node at (-20pt,0pt){$-$};

\draw (0pt,0pt) circle (4.5pt);
\node at (0pt,0pt){$+$};
\draw [thick](4.5-20pt,0pt)--(15.5-20pt,0pt);
\draw (20pt,0pt) circle (4.5pt);
\node at (20pt,0pt){$-$};

\draw (40pt,0pt) circle (4.5pt);
\node at (40pt,0pt){$+$};
\draw [thick](24.5pt,0pt)--(35.5pt,0pt);
\draw (60pt,0pt) circle (4.5pt);
\node at (60pt,0pt){$-$};

\draw (80pt,0pt) circle (4.5pt);
\node at (80pt,0pt){$+$};
\draw [thick](64.5pt,0pt)--(75.5pt,0pt);
\draw (100pt,0pt) circle (4.5pt);
\node at (100pt,0pt){$-$};

\draw (120pt,0pt) circle (4.5pt);
\node at (120pt,0pt){$+$};
\draw [thick](104.5pt,0pt)--(115.5pt,0pt);
\draw (140pt,0pt) circle (4.5pt);
\node at (140pt,0pt){$-$};
\draw [thick](144.5pt,0pt)--(150.5pt,0pt);
   \end{scope}

\begin{scope}[yshift=154pt]
\small
\node at (-56pt,0pt){$\alpha=2\pi$:};


\notsotiny
\draw (-20pt,0pt) circle (4.5pt);
\node at (-20pt,0pt){$+$};

\draw (0pt,0pt) circle (4.5pt);
\node at (0pt,0pt){$-$};

\draw (20pt,0pt) circle (4.5pt);
\node at (20pt,0pt){$+$};


\draw (40pt,0pt) circle (4.5pt);
\node at (40pt,0pt){$-$};

\draw (60pt,0pt) circle (4.5pt);
\node at (60pt,0pt){$+$};

\draw (80pt,0pt) circle (4.5pt);
\node at (80pt,0pt){$-$};

\draw (100pt,0pt) circle (4.5pt);
\node at (100pt,0pt){$+$};

\draw (120pt,0pt) circle (4.5pt);
\node at (120pt,0pt){$-$};

\draw (140pt,0pt) circle (4.5pt);
\node at (140pt,0pt){$+$};

\end{scope}

\end{tikzpicture}
\end{equation}
Similar to \eqref{eq:H_config}, we use ``$+$'' (resp. ``$-$'') to represent a lattice site $p$ with non-zero single-spin term, with the sign representing the corresponding sign of the coefficient of $\sigma^z_p$,
and ``$\bullet$'' represents a lattice site with vanishing single-spin term, as occurs at $w_1 = \cos\alpha=0$.
 The $\mr{U}(1)$ charge operator is defined as $\hat Q=\sum_p \hat Q_p=\sum_p \frac{1}{2}\sigma_p^3$.
 It is straightforward to check that $\hat Q$ commutes with the Hamiltonians $H_\mr{1d}(w)$ for arbitrary $w\in S^1$.
As we tune the parameters $\alpha\in[0,2\pi]$ in this model, a quantized $\mr{U}(1)$ charge will be pumped along the $d=1$ chain.
This quantized charge corresponds to the topological invariant, which is  an integral of 1-form $\tilde Q^{(1)}$:
\begin{equation}
\label{Q_pump}
Q_{\rm pump}=\int_{S^1} \tilde Q^{(1)}(f),
\end{equation}
with
\begin{equation}
\label{Q_tilde}
\tilde Q^{(1)}(f)=\langle Q^{(1)},\delta f\rangle.
\end{equation}
Here $Q^{(1)}$ is the one-form constructed from the $U(1)$ invariant MPS, 
$f(p)=\Theta(p-a)$ where $a\in 2\mathbb Z+1/2$ as indicated by the dashed line in \eqref{H_configU1}, and $\delta f$ is defined as $(\delta f)(p,q)=f(q)-f(p)$.
The explicit expression of $Q^{(1)}$ is
\begin{equation}
\label{T1_U1}
Q^{(1)}_{p_0 p_1}=\dd_{p_0}\langle \hat Q_{p_1}\rangle-\dd_{p_1}\langle \hat Q_{p_0}\rangle.
\end{equation}

The physical meaning of $\tilde Q^{(1)}(f)$ in \eqref{Q_tilde} can be intuitively understood as follows. With the simple choice of $f(p)=\Theta(p-a)$,
\eqref{Q_tilde} can be rewritten as
\be
\label{tildeQ_002}
\tilde Q^{(1)}(f)=\sum_{p<a<q} Q_{pq}^{(1)}=
\dd_L\langle \hat Q_R\rangle-\dd_R\langle \hat Q_L\rangle,
\ee
where we have defined $\dd_L:=\sum_{p<a} \dd_p$, $\dd_R=\sum_{p>a}\dd_p$, 
$\hat Q_L=\sum_{p<a}\hat Q_p$, and $\hat Q_R=\sum_{p>a}\hat Q_p$.
They satisfy the following relation:
\be
\label{dLdRd}
\begin{split}
&\dd_L+\dd_R=\dd=\sum_{p\in \Lambda} \dd_p,\\
&\hat Q_L+\hat Q_R=\hat Q=\sum_{p\in \Lambda}\hat Q_p.
\end{split}
\ee
Then \eqref{tildeQ_002} can be written as:
\be
\begin{split}
\tilde Q^{(1)}
=&(\dd-\dd_R)\langle \hat Q_R\rangle-\dd_R[\langle \hat Q\rangle-\langle \hat Q_R\rangle]\\
=&\dd\langle \hat Q_R\rangle=-\dd\langle \hat Q_L\rangle,
\end{split}
\ee
where we have considered the total charge $\langle \hat Q\rangle$ is conserved.
That is, the closed 1-form $\tilde Q^{(1)}$ measures the increasing of charge in the right half chain, or equivalently the decreasing of charge in the left half chain, as expected.

Now we apply the above formulas to the toy model in \eqref{H_1d_U1}. We will consider both the left and right canonical MPSs.
Since our $1d$  model is composed of decoupled dimers, one can consider the Schmidt decomposition on each dimer.
For the dimer living on sites $p$ and $p+1$, the wavefunction can be expressed as
\begin{equation}
\label{dimer_wf}
|\psi\rangle_{p,p+1}=\sum_{\alpha}\Lambda_{\alpha} |\alpha\rangle_p \otimes |\alpha\rangle_{p+1}.
\end{equation}
The Schmidt coefficients $\Lambda_\alpha$ can always be chosen positive, 
the states $\{|\alpha\rangle_p\}$ and $\{|\alpha\rangle_{p+1}\}$ form orthonormal sets in the Hilbert spaces $\mathcal H_p$
and $\mathcal H_{p+1}$ respectively, i.e., 
$\langle \alpha|\beta\rangle_p=\langle \alpha|\beta\rangle_{p+1}=\delta_{\alpha\beta}$.
By normalization, we have $\sum_\alpha\Lambda_\alpha^2=\langle \psi|\psi\rangle_{p,p+1}=1$.
One can rewrite the wavefunction in \eqref{dimer_wf} as
\begin{equation}
|\psi\rangle_{p,p+1}=
\sum_{s_p}\sum_{s_{p+1}} \Gamma^{s_p} \, \Lambda \, \Gamma^{s_{p+1}}\, |s_p\rangle\otimes |s_{p+1}\rangle,
\end{equation}
where $\Gamma^{s_p}_{0\alpha }=\langle s_p|\alpha\rangle_p$, $\Gamma^{s_{p+1}}_{\alpha 0}=\langle s_{p+1}|\alpha\rangle_{p+1}$, and $s_p,\, s_{p+1}\in \{\uparrow,\,\downarrow\}$.
In the left canonical MPS, one can choose
\begin{equation}
\label{left_canonical_appendix}
A^{s_p}=\Gamma^{s_p}, \quad A^{s_{p+1}}=\Lambda\Gamma^{s_{p+1}}.
\end{equation}
They satisfy the left canonical condition:
\begin{equation}
\begin{split}
\sum_{s_p}(A^{s_p})^\dag A^{s_p}=\mathbb I_2,\quad \sum_{s_p} A^{s_p} (\Lambda)^2 (A^{s_p})^\dag=1,\\
\sum_{s_{p+1}} (A^{s_{p+1}} )^\dag A^{s_{p+1}}=1,\quad
\sum_{s_{p+1}} A^{s_{p+1}} (A^{s_{p+1}})^\dag =(\Lambda)^2. 
\end{split}
\end{equation}
Similarly, in the right canonical MPS, one can choose the tensors $A^{s_p}$ and $A^{s_{p+1}}$ as
\begin{equation}
\label{right_canonical_appendix2}
A^{s_p}=\Gamma^{s_p}\Lambda, \quad A^{s_{p+1}}=\Gamma^{s_{p+1}},
\end{equation}
which satisfy the right canonical conditions
\begin{equation}
\label{Left_canonical_Condition}
\small
\begin{split}
\sum_{s_p} A^{s_p} (A^{s_p})^\dag=1,\quad \sum_{s_p}  (A^{s_p})^\dag A^{s_p}=(\Lambda)^2,\\
\sum_{s_{p+1}} A^{s_{p+1}} (A^{s_{p+1}})^\dag=\mathbb I_2, \quad \sum_{s_{p+1}}
(A^{s_{p+1}})^\dag  (\Lambda)^2
A^{s_{p+1}}=1.
\end{split}
\end{equation}

In our toy model, for the choice of the left canonical form of MPS in \eqref{left_canonical_appendix}, we have
\begin{equation}
\begin{split}
&\Gamma^{\uparrow_p}=(1,0),\quad \Gamma^{\downarrow_p}=(0,-1),\\
&\Gamma^{\uparrow_{p+1}}=(0,1)^T,\quad \Gamma^{\downarrow_{p+1}}=(1,0)^T,
\end{split}
\end{equation}
and
\begin{equation}
\label{lambda2}
\Lambda=\text{diag}(\sqrt{(1+\cos\alpha)/2}, \sqrt{(1-\cos\alpha)/2}).
\end{equation}
Here $p\in 2\mathbb Z+1$ and $0\le \alpha\le \pi$.
Similarly one can obtain the MPS for $\pi< \alpha\le 2\pi$.

Since our lattice model is composed of decoupled dimers, one can find that for the chosen $f(p)$ in \eqref{H_configU1},
the closed 1-form $\tilde Q^{(1)}$ is nonzero only when $0\le\alpha\le \pi$ and it is contributed by 
the tensors at $p$ and $p+1$ as: 
\begin{equation}
\label{Q1_toy}
\tilde Q^{(1)}(f)=Q^{(1)}_{p,p+1}=\dd_p \langle \hat Q_{p+1}\rangle-\dd_{p+1} \langle \hat Q_p\rangle.
\end{equation}
More concretely, in the left canonical form, one has
\begin{equation}
\langle \hat Q_{p}\rangle=\frac{\cos\alpha_{p+1}}{2}, \quad\langle \hat Q_{p+1}\rangle=-\frac{\cos \alpha_{p+1}}{2}.
\end{equation}
Here we use $\alpha_{p+1}$ to emphasize that it parameterizes the tensor at site $p+1$. Then based on \eqref{Q1_toy}, one can obtain the 1-form $\tilde Q^{(1)}(f)$ as
\begin{equation}
\label{Q1_toy2}
\tilde Q^{(1)}(f)=Q^{(1)}_{p,p+1}=\dd_p \langle \hat Q_{p+1}\rangle-\dd_{p+1} \langle \hat Q_p\rangle=\frac{\sin\alpha}{2}d\alpha
\end{equation}
for $0\le \alpha\le \pi$ and $\tilde Q^{(1)}(f)=0$ for $\pi \le \alpha\le 2\pi$. It is reminded that
$
\dd_p:=d\alpha_p\frac{\partial}{\partial \alpha_p},
$
and in the last step of \eqref{Q1_toy} we have taken $\alpha_p=\alpha$. Therefore, the topological invariant is
\begin{equation}
\label{Qinvariant_1}
Q_{\text{pump}}=\int_{S^1} \tilde{Q}^{(1)}(f)=\int_0^\pi \frac{\sin\alpha}{2}d\alpha=1,
\end{equation}
which measures the charge pumped across $x=a$ (See the dashed line in \eqref{H_configU1})
as we tune the parameter adiabatically along $S^1$.

\bigskip
One can also check the choice of the right canonical form of MPS explicitly. One has
\begin{equation}
\langle \hat Q_p\rangle=\frac{\cos\alpha_p}{2},\quad \langle \hat Q_{p+1}\rangle=-\frac{\cos \alpha_p}{2}.
\end{equation}
Here $\alpha_p$ arameterizes the tensor at site $p$. Then based on \eqref{Q1_toy}, one can obtain 
the 1-form $\tilde Q^{(1)}(f)$, which is the same as the result for the left canonical form of MPS.

One can also consider the `symmetric canonical form', by choosing
$A^{s_p}=\Gamma^{s_p}\sqrt{\Lambda},\,A^{s_{p+1}}=\sqrt{\Lambda}\Gamma^{s_{p+1}}$ for $0\le \alpha\le\pi$ and $A^{s_{p-1}}=\Gamma^{s_{p-1}}\sqrt{\Lambda},\,A^{s_{p}}=\sqrt{\Lambda}\Gamma^{s_{p}}$ for $\pi\le \alpha\le 2\pi$ where $p\in 2\mathbb Z+1$. Again, it is found the 1-form $Q^{(1)}(f)$ is the same as that for the left/right canonical form of MPS.

Note that the 1-form $\tilde Q^{(1)}(f)$ depends on the choice of $f$, 
which is similar to the feature of higher Berry curvatures $\Omega^{(3)}(f)$ \cite{2024Sommer}. For example, by choosing $f'(p)=\Theta(p-a)$ with $a\in 2\mathbb Z-1/2$ (i.e., shifting the 
dashed line in \eqref{H_configU1} by an odd number of sites), then for the various canonical forms introduced above, one can obtain
\begin{equation}
\label{Q1_f'}
\tilde Q^{(1)}(f')=-\frac{\sin\alpha}{2}d\alpha, \quad \pi\le \alpha\le 2\pi,
\end{equation}
and $\tilde Q^{(1)}(f')=0$ for $0\le \alpha\le \pi$. However, the topological invariant 
$Q_{\text{pump}}=\int_{S^1} \tilde Q^{(1)}(f')=1$ is the same as \eqref{Qinvariant_1}.

Physically, $\tilde Q^{(1)}(f)$ corresponds to the ``current" which may vary in space for an inhomogeneous system. The integral of this current, which measures the charge pumped along the $1d$ system, is a topological invariant.

\subsection{Higher Thouless charge pump}

Higher Thouless charge pump in a general SRE system with a global $\mr{U}(1)$ symmetry was firstly discussed in Ref.\onlinecite{KS2020_higherthouless}.
For a $d$ dimensional lattice system over $X=S^d$, one can define $d$-form $\tilde Q^{(d)}$. The topological invariant $\int_{S^d} \tilde Q^{(d)}\in \mathbb Z$
characterizes the quantized Thouless charge pump in SRE systems. 
Here, we show our construction of $d$ forms gives the expected topological invariant in systems with nontrivial higher Thouless pumps.
We will be brief since the discussion follows the same procedure based on the bulk-boundary correspondence in Sec.\ref{Sec:higherD}.

Let us give an explicit example of $d=2$ lattice system over $X=S^2$. As before let $(w_1,w_2,w_3)=w\in S^2\subseteq \R^3$ be the standard embedding of the two-sphere in $\R^3$, and let $(\hat{w}_1,\hat{w}_2)$ be the unit vector of $(w_1,w_2)$.
The family of systems can be constructed from the $d=1$ system in \eqref{H_1d_U1} based on the suspension construction as introduced in Ref.\onlinecite{qpump2022}.
The Hamiltonians are 
\begin{equation}
    \label{H_2d_U1}
	H_\mr{2d}(w)=\sum_{x_1\in \mathbb Z} H_{\mr{1d},x_1}((-1)^{x_1}\hat{w}_1,\hat{w}_2)+H^\mr{int}_{x_1,x_1+1}(w)
\end{equation}
where 
$H_{\mr{1d},x}$ is the $d=1$ Hamiltonian from \eqref{H_1d_U1} along the $d=1$ sublattice of points with first coordinate $x$. The coupling is 
\begin{equation}
H^\mr{int}_{x_1,x_1+1}(w)=h_{\vb x}(w_3)\sum_{x_2\in \Z}\bm{\sigma}_{x_1,x_2}\cdot \bm{\sigma}_{x_1+1,x_2}
\end{equation}
For $w_3=0$, the $d=2$ system is a collection of decoupled $d=1$ systems with alternating charges.
One can check that $H_{\mr{2d}}$ in \eqref{H_2d_U1} is gapped everywhere over $S^2$.

The closed 2 forms $\tilde Q^{(2)}(f)$ characterizing the Thouless charge pump are
\begin{equation}
\label{Q2}
\tilde Q^{(2)}(f)=\langle Q^{(2)}, \delta f_1\cup \delta f_2\rangle,
\end{equation}
where $Q^{(2)}$ are (non-closed) 2-forms constructed from $d=2$ PEPSs:
\begin{equation}
\label{T_rpq}
Q^{(2)}_{rpq}=
\dd_r\dd_p\langle \hat Q_q\rangle-\dd_r\dd_q\langle \hat Q_p\rangle
-\dd_q \dd_p\langle \hat Q_r\rangle+\dd_p \dd_q\langle \hat Q_r\rangle
\end{equation}
up to an exact form. $f_{1,2}=\Theta(x_{1,2}(p)-a_{1,2})$ are chosen in the same way as in Sec.\ref{Sec:higherD} (See also Fig.\ref{WF_2d_clutch}).

Instead of calculating $Q^{(2)}(f)$ in an infinite system directly, which is involved, we study it in a semi-infinite system based on bulk-boundary correspondence.
We impose a spatial boundary along $x_1=N$ by retaining only those lattice sites with $x_1\le N$. In our lattice model, $Q_{pqr}^{(2)}$ is a local quantity, and $\tilde Q^{(2)}(f)$
is contributed by the tensors near $(x_1,x_2)=(a_1,a_2)$. As long as we choose $(a_1,a_2)$ deep in the bulk, $\tilde Q^{(2)}(f)$ are the same for the infinite system and the semi-infinite system. In the semi-infinite system, we can write $\tilde Q^{(2)}(f)$ as
\begin{equation}
\tilde Q^{(2)}(f)=\dd \langle Q^{(1)}, \,f_1\cup \delta f_2\rangle=: \dd \omega^{(1)}(f),
\end{equation}
where $Q^{(1)}$ is defined in \eqref{T1_U1}.
For a system with nontrivial higher Thouless pump, $\omega^{(1)}$ is not globally well defined over $X=S^2$.
But if we cover $S^2$ with two charts $D_N^2$ and $D_S^2$, then $\omega^{(2)}$ can be globally well defined over each chart.
More explicitly, we choose $D_N^2$ ($D_S^2$) as the subspace of $S^2$ with $w_3\ge 0$ ($w_3\le 0$).
Over $D_N^2$ ($D_S^2$), the boundary of the $2d$ lattice is imposed along $x_1=N$ ($x_1=N-1$) where $N\in 2\mathbb Z$ (See Fig.\ref{WF_2d_clutch}).
One can find the ground states of $H_{2d}$ are gapped everywhere over $D_N^2$ and $D_S^2$ respectively. 
The 1-forms $\omega_N^{(1)}$ ($\omega_S^{(1)}$) are globally well defined for the corresponding family of PEPSs.
Then one has
\begin{equation}
\small
\int_{S^2}\tilde Q^{(2)}(f)=\int_{D_N^2}\tilde Q^{(2)}+\int_{D_S^2}\tilde Q^{(2)}=\int_{S^1}(\omega_N^{(1)}-\omega_S^{(1)}),
\nonumber
\end{equation}
where $S^1=D_N^2\cap D_S^2$. Note that over $S^1$ where $w_3=0$, $H_{2d}$ in \eqref{H_2d_U1} becomes decoupled $1d$ systems.
In this case, $\omega_N^{(1)}-\omega_S^{(1)}=\tilde Q^{(1)}$, where $\tilde Q^{(1)}$ is the closed 1-form for the decoupled $1d$ system along $x_1=N$.
Therefore, we have 
\begin{equation}
\int_{S^2}\tilde Q^{(2)}=\int_{S^1}\tilde Q^{(1)}=1.
\end{equation}
Therefore, the topological invariant that characterizes the higher Thouless pump in $H_{2d}$ in \eqref{H_2d_U1} is quantized.

We can further study those lattice models obtained from the suspension construction in even higher dimensions\cite{qpump2022} by following the similar procedures above.
Two essential ingredients in this procedure are (i) the descent equations and (ii) the locality property of $Q^{(n)}$. Here $Q^{(n)}$ is related to the closed form $\tilde Q^{(n)}$ through $\tilde Q^{(n)}=\langle Q^{(n)},b\rangle$. 
While the locality property of $Q_{p_0\cdots p_n}^{(n)}$ is apparent in our dimerized lattice models, it is an interesting future problem to show this property for general $U(1)$ invariant gapped ground states.
\let\i\oldi
\bibliography{MPSref.bib}
\end{document}